\documentclass{aa}
\pdfoutput=1
\usepackage{natbib}
\usepackage[varg]{txfonts}
\usepackage{hyperref}
\usepackage[dvipsnames]{xcolor}
\hypersetup{
  colorlinks,
  linkcolor={blue!90!black},
  citecolor={blue!90!black},
  urlcolor={blue}
}
\usepackage[T1]{fontenc}
\usepackage{graphicx}
\usepackage{txfonts}
\usepackage{amsmath}
\usepackage{float}
\usepackage{amssymb}
\usepackage{multirow,bigdelim}
\usepackage{mathtools}
\usepackage{tabulary}
\usepackage{tablefootnote}
\usepackage{threeparttable}
\usepackage[switch]{lineno}
\usepackage{bm}
\usepackage{soul}
\usepackage{comment}
\usepackage{euclid}

\newcommand{\hp}{\texttt{HEALPix}}
\newcommand{\flask}{\texttt{Flask}}
\newcommand{\salmo}{\texttt{Salmo}}
\newcommand{\flasmo}{\texttt{Flask + Salmo}}
\newcommand{\lcdm}{$\Lambda$CDM}
\newcommand{\pcl}{pseudo-$C_{\ell}$}

   \title{\textit{KiDS} \& \textit{Euclid}: Cosmological implications of a pseudo angular power spectrum analysis of KiDS-1000 cosmic shear tomography\thanks{This paper is published on behalf of the KiDS Collaboration and the Euclid Consortium.}}

\newcommand{\orcid}[1]{} 
\author{A.~Loureiro$^{1,2,3}$\thanks{\email{arthur.loureiro@ucl.ac.uk}},
L.~Whittaker$^{1,4}$,
A.~Spurio Mancini$^{1,5}$, 
B.~Joachimi$^{1}$, 
A.~Cuceu$^{1}$,
M.~Asgari$^{6,3}$, 
B.~St\"olzner$^{1}$, 
T.~Tr\"oster$^{3}$, 
A.H.~Wright$^{7}$, 
M.~Bilicki$^{8}$
% ...
}

\institute{Affiliations are listed at the end of the paper.}
   \date{Received October 19, 2021; accepted June 27, 2022}
\begin{document}

  \abstract
   {We present a tomographic weak lensing analysis of the Kilo Degree Survey Data Release 4 (KiDS-1000), using a new pseudo angular power spectrum estimator (\pcl) under development for the ESA Euclid mission. Over 21 million galaxies with shape information are divided into five tomographic redshift bins, ranging from 0.1 to 1.2 in photometric redshift. We measured \pcl{} using eight bands in the multipole range $76<\ell<1500$ for auto- and cross-power spectra between the tomographic bins. A series of tests were carried out to check for systematic contamination from a variety of observational sources including stellar number density, variations in survey depth, and point spread function properties. While some marginal correlations with these systematic tracers were observed, there is no evidence of bias in the cosmological inference. B-mode power spectra are consistent with zero signal, with no significant residual contamination from E/B-mode leakage. We performed a Bayesian analysis of the \pcl{} estimates by forward modelling the effects of the mask. Assuming a spatially flat \lcdm{} cosmology, we constrained the structure growth parameter $S_8 = \sigma_8(\Omega_{\rm m}/0.3)^{1/2} = 0.754_{-0.029}^{+0.027}$. When combining cosmic shear from KiDS-1000 with baryon acoustic oscillation and redshift space distortion data from recent Sloan Digital Sky Survey (SDSS) measurements of luminous red galaxies, as well as the Lyman-$\alpha$ forest and its cross-correlation with quasars, we tightened these constraints to $S_8 = 0.771^{+0.006}_{-0.032}$. These results are in very good agreement with previous KiDS-1000 and SDSS analyses and confirm a $\sim 3\sigma$ tension with early-Universe constraints from cosmic microwave background experiments.}

             
   \titlerunning{Pseudo-$C_{\ell}$ analysis of KiDS-1000 cosmic shear}
   \authorrunning{Loureiro et al.}
   \maketitle
%
\section{Introduction}

Since the turn of the millennium, there has been a dramatic increase in the number and precision of new cosmological observations. Measurements of the cosmic microwave background (CMB) radiation have achieved extraordinary precision for the parameters of the standard cosmological model --- the $\Lambda$CDM model, which contains mostly the cosmological constant ($\Lambda$) and cold dark matter (CDM). The most stringent constraints were obtained by the {Planck} Mission \citep{Planck2018-cosmology}, which demonstrated an unprecedented constraining power, measuring some of these parameters with percent-level precision. The Planck Mission, however, did not contradict the standard model of cosmology. Instead, it established the \lcdm{} model as a difficult paradigm to overturn for other cosmological probes and future surveys \citep{2019-Efstathiou-CamSpec,2020-Efstathiou-LCDM}.

Measurements of the CMB provide insight into the characteristics of the early Universe, with information coming from the surface of last scattering, at redshift $z\approx1100$. If the \lcdm{} paradigm is correct, one expects agreement between the cosmological parameters estimated using CMB experiments and those recovered from late-time probes. In other words, late-time cosmological parameters such as the amplitude of the matter density fluctuations, $\sigma_8$, predicted by CMB observations, should match what we observe today with galaxy surveys. Several studies seem to suggest tensions between late-time probes and the CMB \citep{2018-Mortsell-tensions, 2019-Verde-tensions}. These tensions arise in a few parameters, such as the Hubble constant, when constrained using Cepheids and type Ia Supernovae using distance ladder measurements \citep{2011-Riess-H0, 2016-RiessHubble, 2016-Bernal-H0-Tension, 2016-DiValentino-tensions, 2017-Lin-H0-Tension, 2020-Efstathiou-LockdownH0, 2020-Riess-H0}, and the structure growth parameter, $S_8 = \sigma_8\sqrt{\Omega_{\rm m}/0.3}$, when measured using cosmic shear surveys \citep{2013-Heymans, 2014-MacCrann-tensions, 2017-CHFT-Shahab, 2017-Hildebrandt-KiDS450, 2018-DES-Y1, 2020-Park-tension, 2019-Hikage-HSC, 2020-Lemos-tension, 2020-Asgari-KiDS-DES, 2020-Joudaki-DES-KV450, 2020-Tilman-K1000-Ext, 2020-Asgari-2ptsK1000, 2020-Heymans-KiDS1000-Cosmology, 2021-Amon-DES-Y3-Shear, 2021-Secco-DES-Y3-Shear, 2021-DES-Y3-3x2pt}.

Cosmic shear is the study of the weak gravitational lensing of background galaxies due to the intervening large-scale structure of the Universe. This lensing effect produces correlations in galaxy shapes, and these correlations can be used to determine how the dark and baryonic foreground matter is distributed (see \citealt{bartelmann01,kilbinger15} for comprehensive reviews of cosmic shear). Moreover, by measuring the cosmic shear signal in tomographic redshift bins, we can study how the distribution of matter evolves over time; this enables us to place tight constraints on the structure formation of the Universe and its evolution with redshift \citep{2021-Hall-S8}, as well as the dark energy equation-of-state and other standard cosmological parameters.

As more data are independently collected by current galaxy surveys, the slight tension in $S_8$ between CMB and cosmic shear surveys does not seem to disappear. More specifically, all results coming from the Dark Energy Survey (DES, \citealt{2018-DES-Y1, 2018-Troxel})  and the Hyper Supreme Camera (HSC, \citealt{2019-Hikage-HSC}) indicate a lower value for the amplitude of matter density fluctuations, consistent with the Kilo Degree Survey (KiDS, \citealt{2020-Hidelbrandt-KV450, 2020-Heymans-KiDS1000-Cosmology}), but also consistent with the \textit{Planck} projections at a $ \gtrsim 1.6 \sigma$ level. Combining a subset of these independent surveys, which all use different instruments and observing strategies, the tension with the early Universe probes is exacerbated. In a re-analysis of DES first year data combined with KiDS, \cite{2020-Asgari-KiDS-DES} and \cite{2020-Joudaki-DES-KV450} demonstrate that the tension can be around 3.2$\sigma$ between cosmic shear and CMB measurements from \textit{Planck}. These consistent results from different experiments encourage us to believe that it is unlikely that this tension arises from an observational systematic error; however, there is still room for different explanations \citep{2021-Joachimi}. From a similar perspective, recent results from the Atacama Cosmology Telescope (ACT), a ground-based CMB experiment, observe a similar tension in $S_8$ with cosmic shear probes when combined with other CMB experiments \citep{2020-ACT-1, 2021-ACT-2}. This also suggests that it is unlikely that the tension we currently observe comes from a systematic contamination in the CMB experiments. If this discrepancy between measurements is indeed induced by systematic contamination, it would have to be shared across several independent datasets with widely different properties and analysis choices. Hopes are that experiments planned for the near future will shine a light on this interesting cosmic puzzle as there is still room to improve the precision on cosmic shear experiments.

Future galaxy surveys such as the  Legacy Survey of Space and Time (LSST), carried out at the Vera C. Rubin Observatory \citep{2012-LSST}, the European Space Agency's \textit{Euclid}  Mission \citep{2011Euclid}, and the National Aeronautics and Space Administration's (NASA) Nancy Grace Roman Space Telescope \citep{2015-Spergel-NancyRoman} will strongly rely on cosmic shear as one of their primary probes to understand the large-scale structure of the Universe. These future surveys will obtain cosmological information from billions of galaxy shapes using a plethora of statistical tools including correlation functions, angular power spectra, mass maps, and bi-spectra. Amongst these, the \pcl{} technique is a direct harmonic space approach to estimate the angular power spectra of cosmic shear and galaxy positions \citep{1973-Peebles, 2005-BrownCastroTaylor, 2011-Hikage-PCLMethod, 2018-Asgari-PCL, 2019-Alonso-naMaster, 2020-Nicola-PCL, 2021-Garcia-Garcia-GrowthOfStructure}.

The \pcl{} technique has its strengths and weaknesses in comparison with other two-point (2pt) function estimators \citep{2004-Efstathiou-PCL, 2018-Asgari-PCL, 2020-Nicola-PCL}. Among its shared advantages with other 2pt functions, \pcl{}s have a fast and simple application to data distributed on the sphere, with no need for a flat-sky approximation. The speed to obtain \pcl{} estimates, compared to 2pt correlation functions, is particularly convenient when estimating covariances from simulated catalogues. Furthermore, measuring the 2pt function in harmonic space increases the speed of theory calculations for cosmological inference as most Einstein-Boltzmann equation solvers use the advantages of calculating correlations in harmonic space. Performing the analysis in harmonic space avoids transforming the theory-vector to real space, as it is done for correlation functions \citep{2002-Schneider-Estimators}, or transforming the data-vector into harmonic space, as is the case for band-power analysis \citep{2016-Becker-Bandpowers,2018-vanUitert-Bandpowers}.

Yet, \pcl{} techniques have one liability regarding partial sky observations. Naturally, any realistic galaxy survey will contain masked parts of the sky due to bright foreground objects, survey strategy and other observational artefacts.\footnote{With the increase of extremely bright low-orbit satellite tracks from mega satellite constellations, this problem will affect the masks geometry of future ground-based surveys even more, with significant consequences for cosmology and other areas of astronomy \citep{2020-Gallozzi-StopStarlink2, 2020-Gallozi-StopStarlink1}.} In the presence of a mask, the mask's characteristic scales, from its shapes and holes, will mix nearby modes in harmonic space causing correlations between multipoles to appear. For spin-2 fields, such as the shear field, partial sky observations can also cause ambiguity between E/B-modes. 
The multipole mixing, however, can be fully accounted for by analytically calculating the mixing matrix from the survey's geometry. 

{In this work, our approach consists of forward-modelling the mode mixing effect due to masking in the theory. Taking this approach, we are least prone to numerical instability from inverting and deconvolving the mixing matrix from \pcl{} measurements. Our \pcl{} method is a prototype for one of the default 2pt functions used in \textit{Euclid} (Euclid Collab., {in prep.}). The motivation for the forward-modelling of the mask effects is so that the estimator can achieve the requirements for a \pcl{} analysis in \textit{Euclid} \citep{2011Euclid,2020-EuclidForecastValidation, 2020-Tutusaus-EuclidForecast}. Although binning the data-vector and mixing matrix to perform a more stable deconvolution (as done in \cite{2019-Hikage-HSC} and \cite{2021-TilmanTSz}) poses no challenges for the data set used in this analysis, for the future \textit{Euclid} survey this is challenging. For \textit{Euclid}, we are required to measure angular power spectra in 100 log-spaced band-powers for 10 to 13 redshift tomographic bins \citep{2020-EuclidForecastValidation, 2020-Tutusaus-EuclidForecast}. This would require dealing with a mixing matrix with 2.1 to 3.51 Million dimensions, posing potential numerical inaccuracies which might well exceed the very tight accuracy requirements for the Figure-of-Merit for the equation-of-state of Dark Energy measured by the \textit{Euclid} Mission \citep{2011Euclid}.}

Hence, we apply a \pcl{} estimator currently being developed for the \textit{Euclid Science Ground Segment} to state-of-the-art data from a Stage-III cosmic shear experiment: the public ESO Kilo Degree Survey (KiDS). KiDS is currently at its fourth Data Release (KiDS DR4, \citealt{2019-KiDS-DR4}), having observed around 1000~deg$^2$ of the sky with nine-band matched photometry. \cite{2020-KiDS-1000-ShearCat} outlined the KiDS-1000 weak lensing catalogue production and validation, with redshift calibration and measurements presented in \cite{2020-Hildebrandt-SOM-KiDS1000}. Using three different 2pt statistics, \cite{2020-Asgari-2ptsK1000} performed a cosmic shear analysis of the KiDS-1000 catalogue, including a series of consistency checks. The main KiDS-1000 analysis is finally presented in \cite{2020-Heymans-KiDS1000-Cosmology}, combining it with galaxy clustering from BOSS DR12 \citep{2016-Reid-BOSSDR12, 2017-Sanchez}, to perform a joint ($3\times 2$pt) analysis for a flat \lcdm{} model, following the methodology described by \cite{2020-KiDS1000-Methods}. Finally, constraints on extensions to \lcdm{} using the same $3\times 2$pt methodology and data are presented in \cite{2020-Tilman-K1000-Ext}.

Here, we demonstrate that a \pcl{} estimator performs comparatively accurately to other 2pt estimators, and we show the advantages of forward-modelling effects of the mask through the use of the mixing matrix as part of the theory modelling in Bayesian inference. We also demonstrate the versatility of using this estimator for systematic contamination analysis, calculating the angular power spectra between our data and some observational artefacts and characteristics. Finally, we show the constraining power of combining cosmic shear data with the latest clustering data from the Baryon Oscillations Spectroscopic Survey and its extension (BOSS and eBOSS, respectively; \citealp{2017-Alam,2020-Gil-Marin, 2021-Bautista, 2020-Bourboux, 2020-eBOSSDR16}). We demonstrate that our independent analysis of the KiDS-1000 cosmic shear catalogue exhibits a similar level of tension with early Universe measurements from Planck Legacy \citep{Planck2018-cosmology} and the Atacama Cosmology Telescope Data Release 4 \citep{2020-ACT-1}.

This study is organised as follows: Section \ref{Sec:DataKiDS} summarises the KiDS-1000 shear catalogue and shape measurements. Section \ref{Sec:Methods} explains the methodology for \pcl{} estimators applied to cosmic shear data, the mixing matrix calculation, the measurements performed on the data for E/B-modes and the covariance matrix estimation. Section \ref{Sec:Systematics} explains our systematic contamination null tests using the \pcl{} estimator. Section \ref{Sec:Inference} details our Bayesian inference: theory modelling for cosmic shear, forward-modelling of the mixing matrix, external galaxy clustering data from baryon acoustic oscillations (BAO) and redshift space distortions (RSD) used to complement the cosmic shear data, as well as the priors and the likelihood implemented in the analysis. Section \ref{Sec:Results} outlines our main results from the inferred posteriors with a discussion on tension with cosmic microwave background measurements. Section \ref{Sec:Conclusions} summarises our findings and conclusions. Extra figures with cosmological constraints and nuisance parameters, as well as a complete table with all parameters probed in our analysis can be found in Appendix \ref{Apdx:Extra}. Finally, Appendix \ref{Apdx:BModesSyst} discusses the details of the systematics null tests for B-modes specifically. 

\section{Cosmic shear data from KiDS-1000}\label{Sec:DataKiDS}
Designed as a weak lensing experiment, the Kilo Degree Survey \citep{2015-KiDS,2015-KiDS-Dr1Dr2, 2017-KiDS-DR3} is a public survey by the European Southern Observatory, currently at its 4th Data Release \citep{2019-KiDS-DR4}\footnote{\url{http://kids.strw.leidenuniv.nl/DR4/}}, with an observed area of around $1000$ deg$^2$. Good seeing nights were prioritised to obtain high-resolution images in the \textit{r}-band, resulting in an excellent average seeing of 0.7 arc-seconds. Like in previous analyses \citep{2019-Wright-KV450, 2020-Hidelbrandt-KV450}, KiDS DR4 photometry is processed with \texttt{Theli} \citep{2005-Erben-Theli, 2013-Schirmer-THELI} and \texttt{AstroWise} \citep{2013-Begeman-ASTROWISE} using five-band NIR photometry ($ZYJHK_s$) from VIKING\footnote{The VISTA Kilo-degree Infrared Galaxy (VIKING) Survey} \citep{2013-Edge-VIKING} combined with four-band optical photometry ($ugri$) from KiDS to estimate photometric redshifts in an accurate and precise way. In this section, we briefly describe the most important and relevant details of the catalogue used in our study.

We perform our main analysis using the latest cosmic shear catalogue from the 4th Data Release (KiDS-1000)\footnote{\url{http://kids.strw.leidenuniv.nl/DR4/lensing.php}}, covering an effective unmasked area of 777.4 deg$^2$ and containing a total of 21\,262\,011 lensed galaxies. Detailed descriptions of the catalogue's construction, including the \texttt{LensFit} \citep{2007-Miller-LensFit, 2008-Kitching-LensFit, 2013-Miller-Lensfit} shape measurement procedures, inverse variance weight estimation, with contributions from measurement error and intrinsic shape noise, as well as systematic contamination tests, can be found in \cite{2020-KiDS-1000-ShearCat}. 
In \cite{2020-KiDS-1000-ShearCat},  several potential contaminants to the data were analysed, including, but not limited to, point spread function (PSF) contamination and the effects of multiplicative and additive bias calibration, finding an impact smaller than $0.1\sigma$ for $S_8 = \sigma_8\sqrt{\Omega_{\rm m}/0.3}$ after the calibration is applied. We perform additional checks in the pseudo-$C_{\ell}$ approach in Sect. \ref{Sec:Systematics}.

\begin{figure}
   \centering
   \includegraphics[width=\columnwidth]{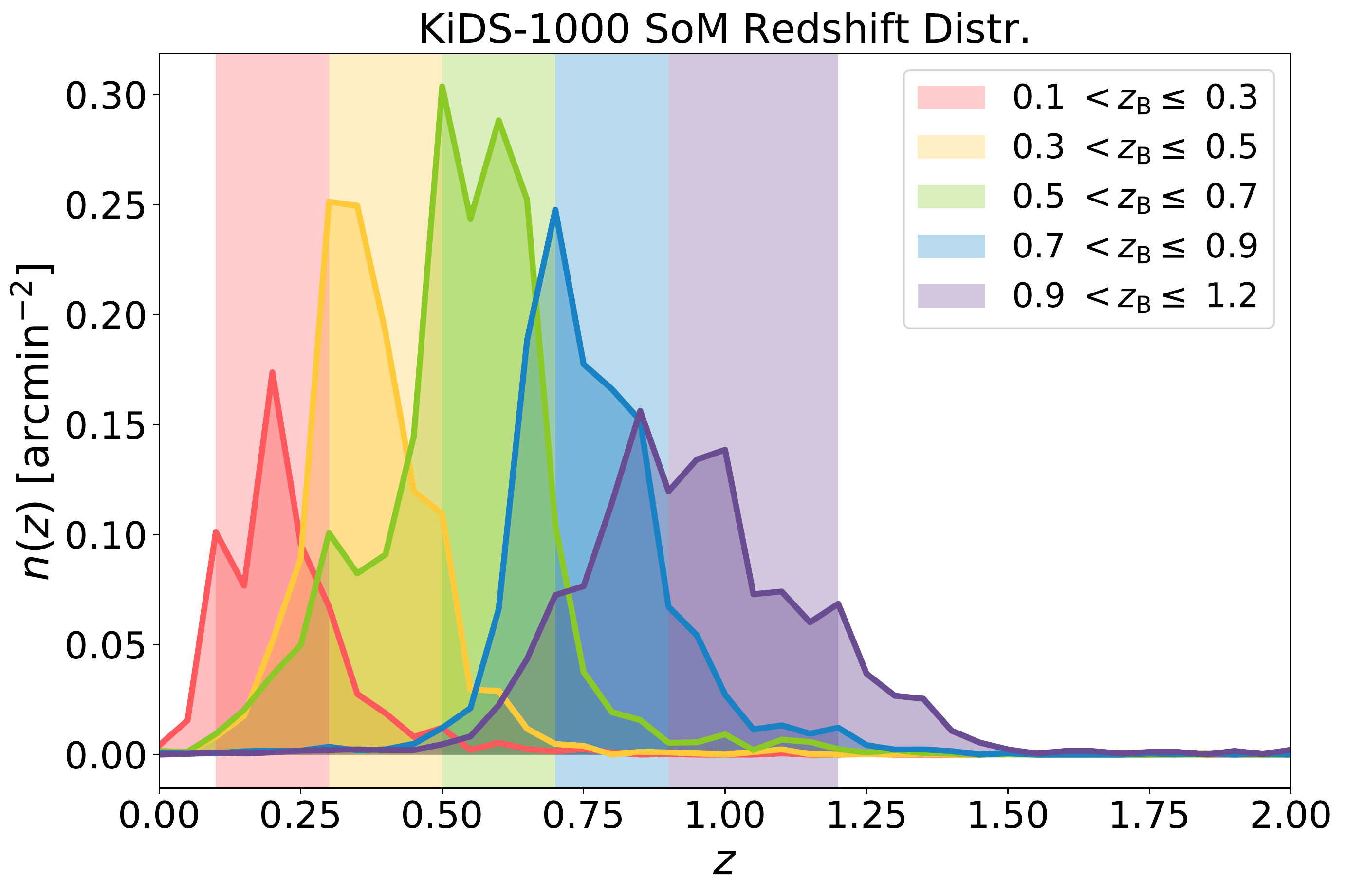}
      \caption{KiDS-1000 redshift distribution for the five tomographic bins used. Solid lines show the redshift distributions obtained via the SOM technique \citep{2020-Hildebrandt-SOM-KiDS1000}. The shaded bands show the corresponding ranges of photometric redshift point estimates, $z_{\rm B}$.}
         \label{Fig:Nz}
\end{figure}

Here, for ease of comparison, we apply the same redshift tomographic binning as in \cite{2020-Asgari-2ptsK1000}, adopting five bins selected using the most probable redshift assigned by \texttt{BPZ} \citep{2000-BPZ,2006-Coe-BPZ, 2014-Raichoor-BPZ}, $z_{\rm B}$. The first four bins have $\Delta z_{\rm B}^{(1\text{-}4)} = 0.2$, ranging from $0.1 < z_{\rm B} \leq 0.9$, while the last tomographic bin ranges from $0.9 < z_{\rm B} < 1.2 $, that is $\Delta z_{\rm B}^{(5)} = 0.3$. The photometric redshift distributions for the lensing sample,  detailed in \cite{2020-Hildebrandt-SOM-KiDS1000}, were estimated using the Self-organising Maps technique (SOM; \citealt{1995-SOM, 1997-Naim-SOM, 2015-Masters-SOM-Euclid, 2019-Kitching-SOM, 2020-Wright-SOM}) to match galaxies using the nine-band photometry to groups with similar properties within spectroscopic samples. We select galaxies detected in all nine bands which have SOM matches and follow extra criteria according to the Gold Sample presented in \cite{2020-Hildebrandt-SOM-KiDS1000}. Figure \ref{Fig:Nz} shows the tomographic selection and the SOM redshift distributions which have been validated with a cross-clustering analysis using spectroscopic surveys (see \citealt{2020-vandenBusch-clustering-photoz} and \citealt{2020-Hildebrandt-SOM-KiDS1000} for details). Table \ref{table:1} contains more details about the samples' redshift tomographic bins and their multiplicative shear calibration corrections.

Cosmic shear, ${\gamma}$, can be estimated using the galaxy ellipticity measurements, corrected by the additive bias, $\epsilon^{\rm corr}$, obtained from the observed galaxy ellipticities, $\epsilon^{\text{obs}}$. In the weak lensing regime ($|{\gamma}| \ll 1$), $\epsilon^{\rm corr}$ is estimated by taking into account the additive bias $ c$,

\begin{align}
    \epsilon^{\rm corr} = {\epsilon^{\text{obs}} - {c}}\, ,
    \label{Eq:Ellips}
\end{align}
where the additive bias, $ {c} = \langle \epsilon^{\text{obs}} \rangle$, is assumed to be the weighted average observed ellipticity over all galaxies in a given tomographic bin. We note that all these quantities (shear, ellipticity and additive bias) are complex quantities, where $\epsilon = \epsilon_1 + \rm{i}\epsilon_2$.

\section{Methodology}\label{Sec:Methods}
This section briefly discusses the pseudo-$C_{\ell}$ methodology used to estimate the cosmic shear angular power spectra. The prototype estimator applied here is currently under development for \textit{Euclid} \citep{2011Euclid}. A detailed study of the methodology performance for Stage IV surveys will be presented in future work. Apart from the noise power spectra estimation, our method is similar to previous works for CMB polarisation power spectra \citep{2003-Kogut-WMAP-PCL, 2005-BrownCastroTaylor} and the full-sky formalism presented by \cite{2011-Hikage-PCLMethod}. 

{In contrast to the approach taken by \cite{2019-Hikage-HSC} and \cite{2021-TilmanTSz}, we focus on forward-modelling the effects of the mask via the {mixing matrix} instead of deconvolving it from the data as done in widely used \pcl{} implementations such as \cite{2002-Hivon-MASTER} and \cite{2019-Alonso-naMaster}. It is important to note that although not obvious at first, the forward modelling of the mask does not imply any extra numerical overhead when modelling the theory vector (discussed in detail in Sect.~\ref{Sec:Theory}). The binning and mixing matrix deconvolution operations do not commute. Therefore, if one chooses to take the standard approach of deconvolving the binned data-vector, similar operations need to be performed for the theoretical modelling during the likelihood analysis (convolve the theory with the mixing matrix, bin the theory-vector, bin the mixing matrix, and finally deconvolve it from the theory vector). All these operations can be written as a single matrix multiplication \citep{2019-Alonso-naMaster}, which is precisely what the forward modelling of the mixing matrix requires. Hence, both approaches have similar computational time with the sole difference that for future galaxy surveys the binned mixing matrix will be much larger and harder to invert numerically. The forward model approach we take in this work avoids this potential issue.}
 
Details on the mixing matrix estimation are given in Sect. \ref{Sec:MixMat}, while modelling of the data-vector is detailed in Sect. \ref{Sec:Theory}. The methodology we use for covariance matrix estimation using simulations is detailed in Sect. \ref{Sec:CovMat}.


\subsection{Pseudo angular power spectrum estimator for cosmic shear}\label{Sec:PCL}
We represent the cosmic shear fields by pixelated maps created from shear estimates derived from the catalogue using a weighted average and correcting for the multiplicative bias, $m_{ z}$, where \textit{z} is the redshift tomographic bin index \citep{2019-Kitching-Bias1, 2020-Kitching-bias2}. The average shear in a pixel localised by the unit vector $\hat{\mathbf{n}}$ is given as

\begin{equation}
    {\gamma}\left(\hat{\mathbf{n}}\right)= \frac{1}{(1+m_{z})} \frac{\sum_{i\in\hat{\mathbf{n}}} w_i \epsilon^{\rm corr}_i}{\sum_{i\in\hat{\mathbf{n}}} w_i}\, ,
    \label{Eq:GammaEst}
\end{equation}
where the summation is taken over all galaxies in the pixel. For each galaxy, $ \epsilon^{\rm corr}_i$ is constructed using Eq. (\ref{Eq:Ellips}), with weights, $w_i$, provided by shape measurement techniques such as \texttt{LensFit} \citep{2007-Miller-LensFit, 2008-Kitching-LensFit, 2013-Miller-Lensfit}. While the additive bias from Eq. \eqref{Eq:Ellips} can be estimated from the catalogue itself, the multiplicative bias ($m_{z}$, \textit{m}-correction) needs to be estimated from careful calibration using pixel-level simulations and is chosen in KiDS to be the average value for all galaxies in a given redshift tomographic bin ${z}$. These corrections are averaged for both shear components and, therefore, are applied to both. The \textit{m}-correction estimation is detailed in \cite{2019-Kannawadi-Calibration} and \cite{2020-KiDS-1000-ShearCat}.

The shear fields can be expressed in terms of a spin-2 field via the E- and B-modes of spherical harmonic decomposition; the observed shear components are analogous to the $Q$ and $U$ Stokes parameters, sharing a similar mathematical structure \citep{1997-Seljak-Stokes,bartelmann01}. For an ideal case with a complete coverage of the sky, the shear field decomposition into spin-2 spherical harmonics, $_{\pm 2}Y_{\ell m}$,is defined as

\begin{align}
    \gamma(\hat{\mathbf{n}}) &= \gamma_1(\hat{\mathbf{n}}) + {\rm i}\gamma_2(\hat{\mathbf{n}}) = \sum_{\ell m}\left({E}_{\ell {m}}+ {\rm i}{B}_{\ell {m}} \right)\,_{+ 2}{Y}_{\ell {m}}(\hat{\mathbf{n}})\, ;\\
    \gamma^*(\hat{\mathbf{n}}) &= \gamma_1(\hat{\mathbf{n}}) - {\rm i}\gamma_2(\hat{\mathbf{n}}) = \sum_{\ell {m}}\left({E}_{\ell {m}} - {\rm i}{B}_{\ell {m}} \right)\,_{- 2}{Y}_{\ell {m}}(\hat{\mathbf{n}})\, , 
\end{align}
where $\gamma(\hat{\mathbf{n}})$ is a complex number and $^*$ represents its complex conjugate. The spherical harmonic coefficients for each mode, over a region $\Omega_{\hat{\mathbf{n}}}$ of the sky, are

\begin{align}
    E_{\ell m}  = & \frac{1}{2}\int {\rm d}\Omega_{\hat{\mathbf{n}}}  \left\{\gamma(\bm\hat{\mathbf{n}})\,_{+2}Y^*_{\ell m} + \gamma^*(\hat{\mathbf{n}})\,_{-2}Y^*_{\ell m}\right\} \, ; \label{Eq:Elm}\\
B_{\ell m} = & \frac{-\text{i}}{2}\int  {\rm d}\Omega_{\hat{\mathbf{n}}} \left\{ \gamma(\bm\hat{\mathbf{n}})\,_{+2}Y^*_{\ell m} - \gamma^*(\bm\hat{\mathbf{n}})\,_{-2}Y^*_{\ell m}\right\}\, . \label{Eq:Blm}
\end{align}

Considering the weak lensing limit, the shear field that emerges from a scalar gravitational field should be a gradient or a curl-free field, containing $B_{\ell m} = 0$ \citep{bartelmann01,2011-Hikage-PCLMethod, kilbinger15}. Although for multiple lenses a B-mode power spectrum can be generated, its power is expected to be orders of magnitude smaller than the E-mode power spectrum, meaning that effectively all the cosmological information should come from the E-mode power spectra and that any significant signals in B-modes would arise from systematic contamination. 

The situation is different for realistic cases when sky coverage is partial. The survey area is now constrained to a region $\mathcal{W}(\hat{\mathbf{n}})$; where $\mathcal{W}(\hat{\mathbf{n}}) = 0$ for regions that are not observed, masked out due to bright stars or bad seeing, or contain no galaxies; and 1 otherwise -- where there are galaxies in the given pixel. In other words, $\mathcal{W}(\hat{\mathbf{n}})$ is an effective binary mask constructed from the catalogue (which already excludes bad regions as detailed in \citealt{2020-KiDS-1000-ShearCat}). Although not the case for the KiDS-1000 dataset, future applications to \textit{Euclid} data $\mathcal{W}(\hat{\mathbf{n}})$ 
need careful consideration on how to deal with a binary mask and shear weights for high-resolution maps (Euclid Collab. et al., \textit{in prep}). 

For partial sky observations on the celestial sphere, the decomposition of the observed shear field into spin-2 spherical harmonics is given by \citep{2005-BrownCastroTaylor}

\begin{align}
    \tilde{\gamma}_1(\hat{\mathbf{n}}) \pm \rm{i}\tilde{\gamma}_2(\hat{\mathbf{n}}) & \equiv \mathcal{W}(\hat{\mathbf{n}})\left[\gamma_1(\hat{\mathbf{n}}) \pm 
    {\rm i}\gamma_2(\hat{\mathbf{n}})) \right] \, \nonumber \\
    & = \sum_{\ell m}\left(\tilde{E}_{\ell m}\pm {\rm i}\tilde{\textit{B}}_{\ell {m}} \right)\,_{\pm 2}Y_{\ell {m}}(\hat{\mathbf{n}})\, ,\label{Eq:GammaTransf}
\end{align}
where we denote the partial sky quantities with a tilde. Here, the pseudo spherical harmonic coefficients, $\tilde{E}_{\ell m}$ and $\tilde{B}_{\ell m}$, are related to the full-sky E- and B-modes as 

\begin{align}
    \tilde{E}_{\ell m} & = \sum_{\ell' m'}\left(E_{\ell m} \tens{W}^+_{\ell\ell' m m'} + B_{\ell m} \tens{W}^-_{\ell\ell' m m'} \right ) \, ; \label{Eq:PseudoE}\\
    \tilde{B}_{\ell m} & = \sum_{\ell' m'}\left(B_{\ell m} \tens{W}^+_{\ell\ell' m m'} - E_{\ell m} \tens{W}^-_{\ell\ell' m m'} \right )\, , \label{Eq:PseudoB}
\end{align}
where the mixing kernels $\tens{W}^{\pm}_{\ell\ell' m m'}$ are explained in more detail in the following section.

Now, considering \textit{i} and \textit{j} as redshift tomographic bins, we can define an estimator for the pseudo angular power spectra for cosmic shear as

\begin{align}
    \tilde{C}^{E_i E_j}_{\ell} & \approx \frac{1}{2\ell + 1} \sum_{m}  \tilde{E}^i_{\ell m} \tilde{E}^j_{\ell m}  - \left\langle\widetilde{\mathcal{N}}^{i,j}_{\ell}\right\rangle\delta_{ij} \, ,
    \label{Eq:PCL_EMode}
\end{align}
with a similar expression for the pseudo B-modes. The last element on the right-hand side of Eq. \eqref{Eq:PCL_EMode} is the average shape noise pseudo power spectrum defined in the following paragraph. 

We estimate the noise power spectrum by randomising the orientation of the galaxy shapes in the catalogue by a random angle $\theta^R$, while keeping the galaxy positions and weights fixed:

\begin{align}
    \epsilon_1^R & = \epsilon^{\text{corr}}_1\cos(\theta^R) - \epsilon^{\text{corr}}_2\sin(\theta^R)\, , \label{Eq:NoiseEps1}\\
    \epsilon_2^R & = \epsilon^{\text{corr}}_2\cos(\theta^R) + \epsilon^{\text{corr}}_1\sin(\theta^R)\, .\label{Eq:NoiseEps2}
\end{align}
This keeps the mask and effective number density of galaxies fixed. Then, we measured the pseudo E/B-mode angular power spectrum of the randomised ellipticities catalogue. This step is repeated 100 times and the mean noise power spectrum is finally subtracted from the data's \pcl{}. Using Eq. \eqref{Eq:NoiseEps1} and \eqref{Eq:NoiseEps2} together with Eq. \eqref{Eq:GammaEst} to define the randomised shear estimates, $\gamma^R$, one can propagate this quantity through Equations \eqref{Eq:GammaTransf} and \eqref{Eq:PseudoE} to obtain $\tilde{E}^{R,i}_{\ell m}$. Finally, the $\tilde{\mathcal{N}}_{\ell}$ term from Eq. \eqref{Eq:PCL_EMode} is simply defined as

\begin{align}
    \tilde{\mathcal{N}}^{i,j}_{\ell} = \frac{1}{2\ell + 1} \sum_{m}  \tilde{E}^{R,i}_{\ell m} \tilde{E}^{R,j}_{\ell m}\, .
\end{align}
It is important to note that even if the number of catalogue randomisations has a very small impact on the measured angular power spectrum, it can have a significant impact on the covariance estimation (cf. Sect.~\ref{Sec:CovMat}).

Although this has been the standard methodology for noise estimates in harmonic space for cosmic shear surveys \citep{2016-Becker-DES-SV-Shear, 2019-Hikage-HSC}, it could potentially lead to a biased estimate of $\tilde{\mathcal{N}}_{\ell}$. Starting from Eq. \eqref{Eq:Ellips} and assuming both the intrinsic ellipticity and shear are Gaussian distributed, by randomly orienting galaxies in a shear catalogue one obtains an estimate of $\sigma^2_{\epsilon_{\rm obs}} = \sigma^2_{\gamma} + \sigma^2_{\epsilon}$, where the shear field variance, $\sigma^2_{\gamma}$, is mixed with the variance from the intrinsic shape and measurement errors, $\sigma^2_{\epsilon}$. However, using the mocks described in Sect. \ref{Sec:CovMat}, we have verified that this effect is negligible for the KiDS-1000 cosmic shear catalogue. This test consisted comparing the estimated noise from ellipticities and pure shear dispersion from the mocks using the method described above. This randomised shape catalogue approach for noise power spectra estimation also proved to be a potential bottleneck for future \textit{Euclid} applications; alternatives such as taking the expectation value of the noise will be implemented for Stage IV surveys.


\subsection{The mixing matrix}\label{Sec:MixMat}
\begin{figure}
   \centering
   \includegraphics[width=\columnwidth]{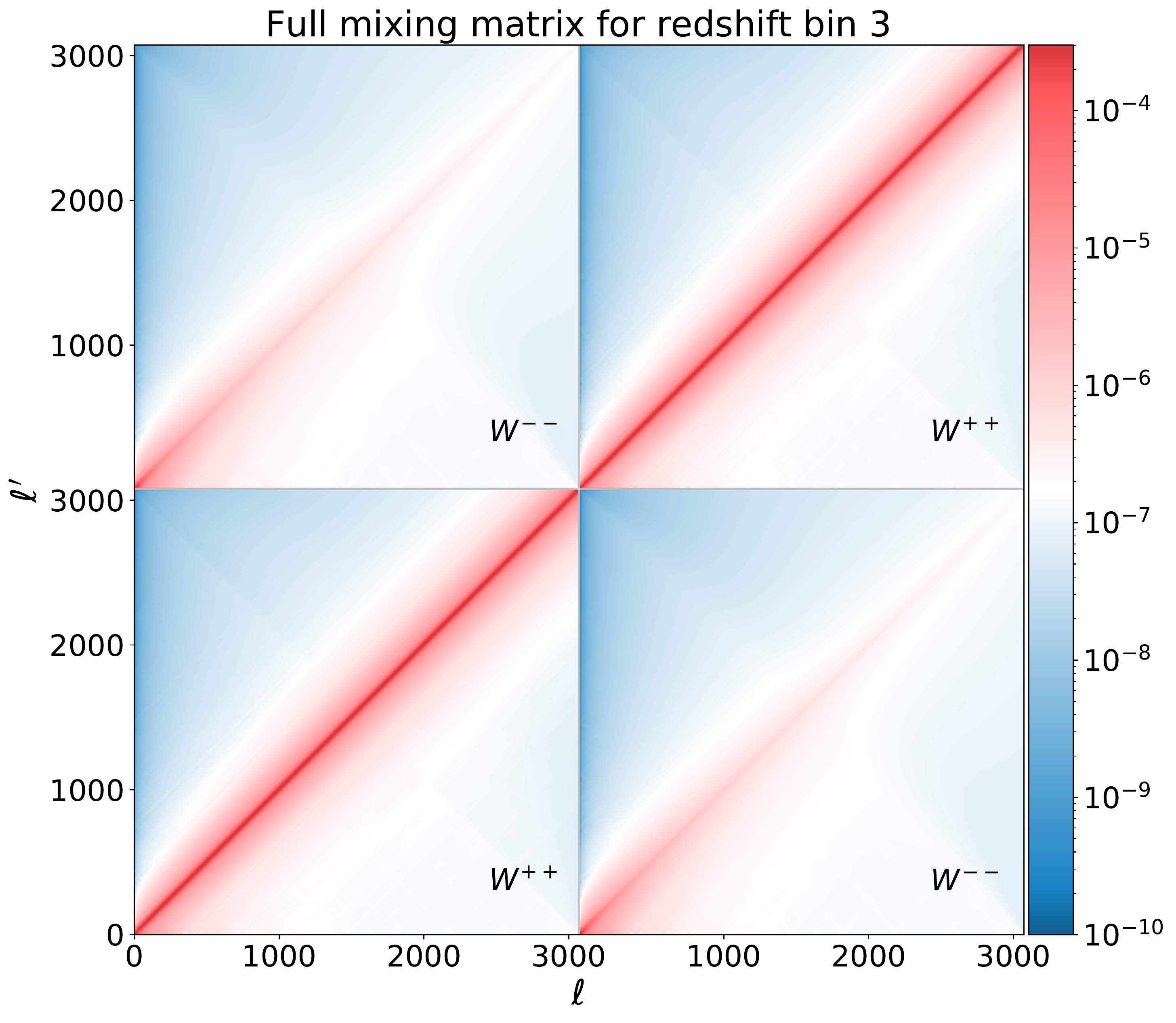}
      \caption{KiDS-1000 full mixing matrix, $\tens{M}_{\ell \ell'}$ from Eqs. \eqref{Eq:ConvCls} and \eqref{Eq:MixMatCls}, for the third redshift tomographic bin. For this case, $\tens{W}^{++}$ in the diagonal is responsible for most of the mixing for the E-modes. In a scenario with non-zero but small B-modes, the off-diagonal terms are still subdominant to the E-mode signal. For each individual block, $\tens{W}^{++}$ and $\tens{W}^{--}$, we show the matrices calculated in the range $2\leq \ell \leq 3070$. From Eqs. \eqref{Eq:MixMat_WPlusPlus} and \eqref{Eq:MixMat_ellellmm} one can see that these objects are not symmetric in $\ell$-$\ell'$.}
         \label{Fig:MixMatBin3}
\end{figure}

The mixing of modes, introduced in Eqs. \eqref{Eq:PseudoE} and \eqref{Eq:PseudoB} via the $\tens{W}^{\pm}_{\ell\ell' m m'}$ terms, is a purely geometrical effect caused by the empty pixels where data are not observed. Differently from galaxy clustering analyses, in cosmic shear the absence of data are taken as the absence of information -- if we do not observe lensed galaxies in a region of the sky, we are unable to estimate shear information for that region. Therefore, to deal with the mode mixing caused by the effective geometry of the survey, we construct effective tomographic masks from the KiDS-1000 cosmic shear catalogue. As previously mentioned, for each redshift tomographic bin this mask, $\mathcal{W}(\hat{\mathbf{n}})$, will be 1 if there are galaxies assigned to that pixel or zero in case the pixel is empty. In other words, in our case the mask is dependent only on the empty pixels for a given tomographic bin.

For a given mask, the expectation value of the pseudo-$C_{\ell}$ estimates taken over all realisations of noise and cosmic variance can be related to the underlying full-sky angular power spectra through the mixing matrix $\tens{M}_{\ell \ell'}$, which contains information about the survey geometry in harmonic space \citep{1973-Peebles, 2005-BrownCastroTaylor, 2011-Hikage-PCLMethod}:

\begin{align}
    \left\langle\tilde{\bm{C}}_{\ell}^{ij}\right\rangle = \sum_{\ell'} \tens{M}_{\ell \ell'}^{ij} \bm{C}_{\ell'}^{ij}\, ,
    \label{Eq:ConvCls}
\end{align}
or, assuming no parity violation modes are present $\left(\tilde{C}_{\ell}^{E_iB_j} = \tilde{C}_{\ell}^{B_iE_j}\right)$,

\begin{gather}
 \begin{pmatrix} \left\langle\tilde{C}_{\ell}^{E_iE_j}\right\rangle  \\ \left\langle\tilde{C}_{\ell}^{B_iB_j}\right\rangle \end{pmatrix}
 =
 \sum_{\ell'}
  \begin{pmatrix}
   \tens{W}^{++}_{\ell\ell'}(i,j) & \tens{W}^{--}_{\ell\ell'}(i,j) \\
   \tens{W}^{--}_{\ell\ell'}(i,j) & \tens{W}^{++}_{\ell\ell'}(i,j) \\
   \end{pmatrix}
   \begin{pmatrix} {C}_{\ell'}^{E_iE_j} \\ {C}_{\ell'}^{B_iB_j}\end{pmatrix}\, .
   \label{Eq:MixMatCls}
\end{gather}
The individual elements of the mixing matrix in the equation above are composed of 

\begin{align}
    \tens{W}^{\pm\pm}_{\ell\ell'}(i,j) = \frac{1}{2\ell+1}\sum_{mm'}\tens{W}^{\pm,i}_{\ell\ell' m m'}\left(\tens{W}^{\pm,j}_{\ell\ell' m m'}\right)^*\, ,
    \label{Eq:MixMat_WPlusPlus}
\end{align}
where $\tens{W}^{\pm}_{\ell\ell' m m'}$ are the same mixing kernels present in Eq. \eqref{Eq:PseudoE} and \eqref{Eq:PseudoB}. These can be defined for each tomographic bin as \citep{2011-Hikage-PCLMethod}

\begin{align}
    \tens{W}^{\pm,i}_{\ell\ell' m m'}  \equiv & \int {\rm d}\Omega_{\hat{\mathbf{n}}} \,_{\pm 2}Y_{\ell' m'} (\hat{\mathbf{n}}) \mathcal{W}^{i}(\hat{\mathbf{n}}) \,_{\pm 2}Y^*_{\ell m} (\hat{\mathbf{n}}) \, ;\nonumber \\
     = & \sum_{\ell''m''}(-1)^m\sqrt{\frac{(2\ell+1)(2\ell'+1)(2\ell'' +1)}{4\pi}}\mathcal{W}^{i}_{\ell''m''} \nonumber \\
    & \times \begin{pmatrix} \ell & \ell' & \ell'' \\ \pm 2 & \mp 2 & 0 \end{pmatrix} \begin{pmatrix} \ell & \ell' & \ell'' \\ m & m' & m'' \end{pmatrix}\, , \label{Eq:MixMat_ellellmm}
\end{align}
where the $2\times 3$ objects above are the Wigner $3j$ symbols, calculated using \texttt{UCLWig3j} library\footnote{\url{https://github.com/LorneWhiteway/UCLWig3j} -- this library optimises the calculation of Wigner $3j$ symbols using the recurrence relation by \cite{SCHULTEN1976269}.}, and

\begin{align}
    \mathcal{W}_{\ell m}^{i} = \oint {\rm d}\Omega_{\hat{\mathbf{n}}} \mathcal{W}^{i}(\hat{\mathbf{n}}) Y^*_{\ell m} (\hat{\mathbf{n}})\, 
\end{align}
is the spin-0 spherical harmonic decomposition of the effective binary mask with the corresponding spin-0 spherical harmonic basis, $Y_{\ell m}(\hat{\mathbf{n}})$. The mode mixing effect introduced by the mask is then forward-modelled into the theory data-vector, with more details presented in Sect. \ref{Sec:Theory}. An example of the full mixing matrix for the third redshift tomographic bin is presented in Fig. \ref{Fig:MixMatBin3}.

Here we would like to point out two important details regarding using a binary mask as opposed to a mask containing the shear weights or a variable depth mask. First, a mask containing shear weights, used as $\mathcal{W}(\hat{\mathbf{n}})$, would cause an imprint from the clustering of galaxies into the mixing matrix, causing an artificial excess power for $\mathcal{W}_{\ell m}$ and biasing the forward-modelling of the mixing matrix into the likelihood (see Sect. \ref{Sec:Theory}). Although this is not a problem for KiDS-1000 given the multipole range, \hp{} resolution, and effective galaxy density ($n_{\rm eff}$), it could be a problem for Stage-IV surveys such as \textit{Euclid}. The pixelisation scheme smooths out the weights information if the resolution is too low for a denser survey. Meanwhile, increasing the resolution too much can cause many pixels containing a single galaxy only, which, given Eq. \eqref{Eq:GammaEst}, would cause the weights information to be lost and bias the analysis. Details regarding how to implement the weights information without causing the aforementioned bias will be investigated in forthcoming work.

When considering variable depth effects via a non-binary mask, one should carefully avoid imprinting clustering information into the mask. In this case, it is not clear how one can include this effect into the estimator or the data-vector using a non-binary mask. Ideally, to include variable depth effects, one should take an approach similar to \cite{2020-Heydenreich} and forward model the effect into the likelihood. Yet, \cite{2020-Heydenreich} shows that this effect has very little impact on the inferred cosmological parameters for a KiDS-like survey. Although we do not include this effect in the estimator or in the mixing matrix calculation, variable depth effects are modelled into the covariance via the \textit{Egretta} mocks (see Sect. \ref{Sec:CovMat}), where their impact is more significant \citep{2020-KiDS1000-Methods}. Therefore, the additional scatter introduced by variable depth is propagated through the inference by including them directly in the covariance matrix.

\subsection{Pseudo-\texorpdfstring{$C_{\ell}$}{Lg} measurements}
\begin{figure}
   \centering
   \includegraphics[width=1\columnwidth]{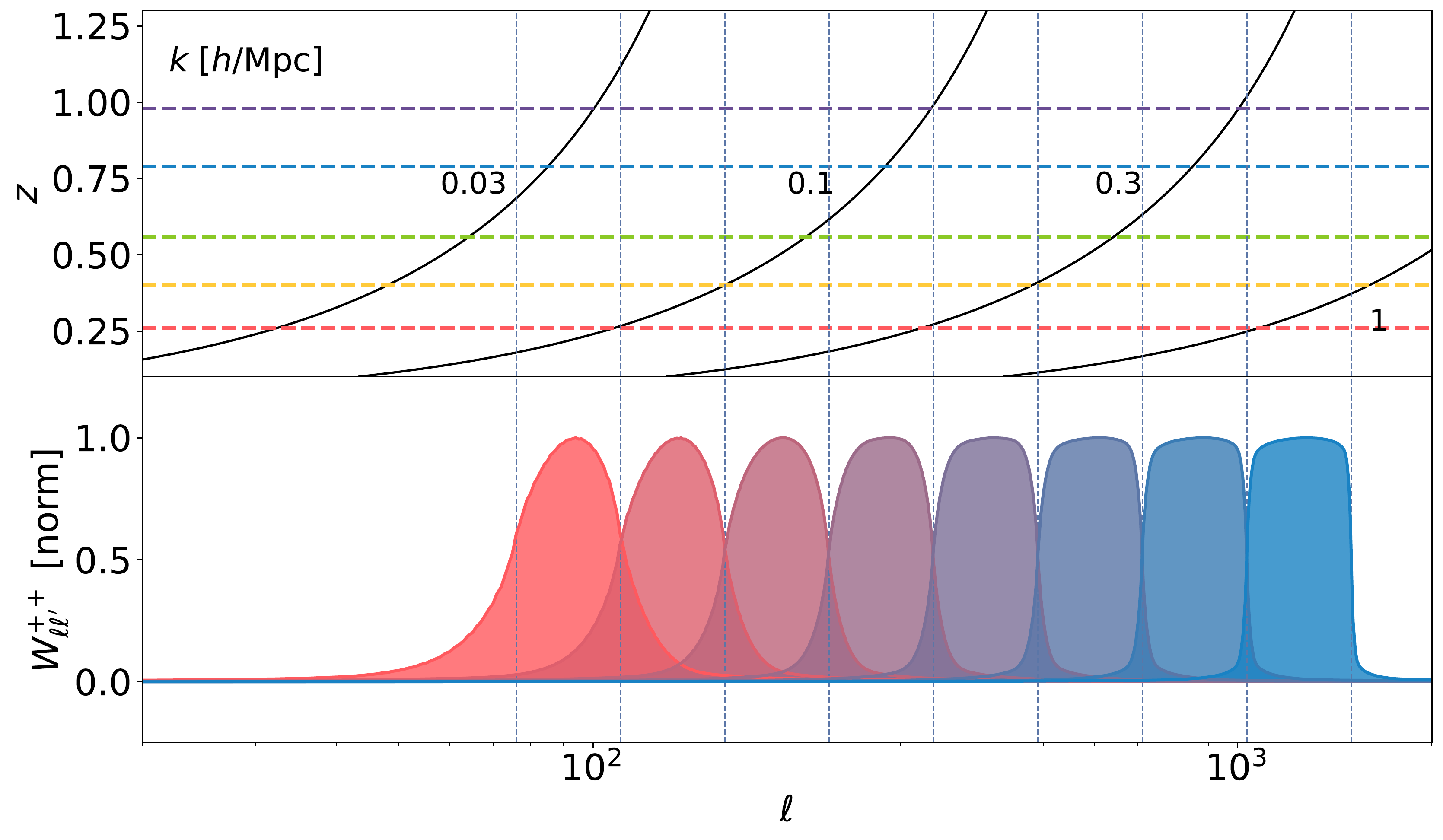}
      \caption{Binned mixing matrices as band-pass filters and the physical scales mixed for each redshift tomographic bin edge. Top: redshift as a function of multipole for different wave-numbers. The coloured horizontal dashed-lines are the centre of the redshift tomographic bins used in the analysis and shown in Fig. \ref{Fig:Nz}; the black lines are curves of constant wavenumber, $k=\ell/\chi(z)$, in units of $h$ Mpc$^{-1}$ for a given $\ell$ and $z$, where $\chi(z)$ is the co-moving distance. Bottom: the KiDS-1000 pseudo-$C_{\ell}$s binned mixing matrices (see Fig. \ref{Fig:MixMatBin3}, for example). These are convolved with the theory $C_{\ell}$s (see Eq.~\ref{Eq:ClsTheory}) to model the effect of multipole mixing introduced by the survey mask. The filters are calculated using all the multipoles inside the eight log-spaced bandwidths from $76 \leq \ell \leq 1500$.}
         \label{Fig:Windows}
\end{figure}

\begin{figure*}
   \centering
   \includegraphics[width=2\columnwidth]{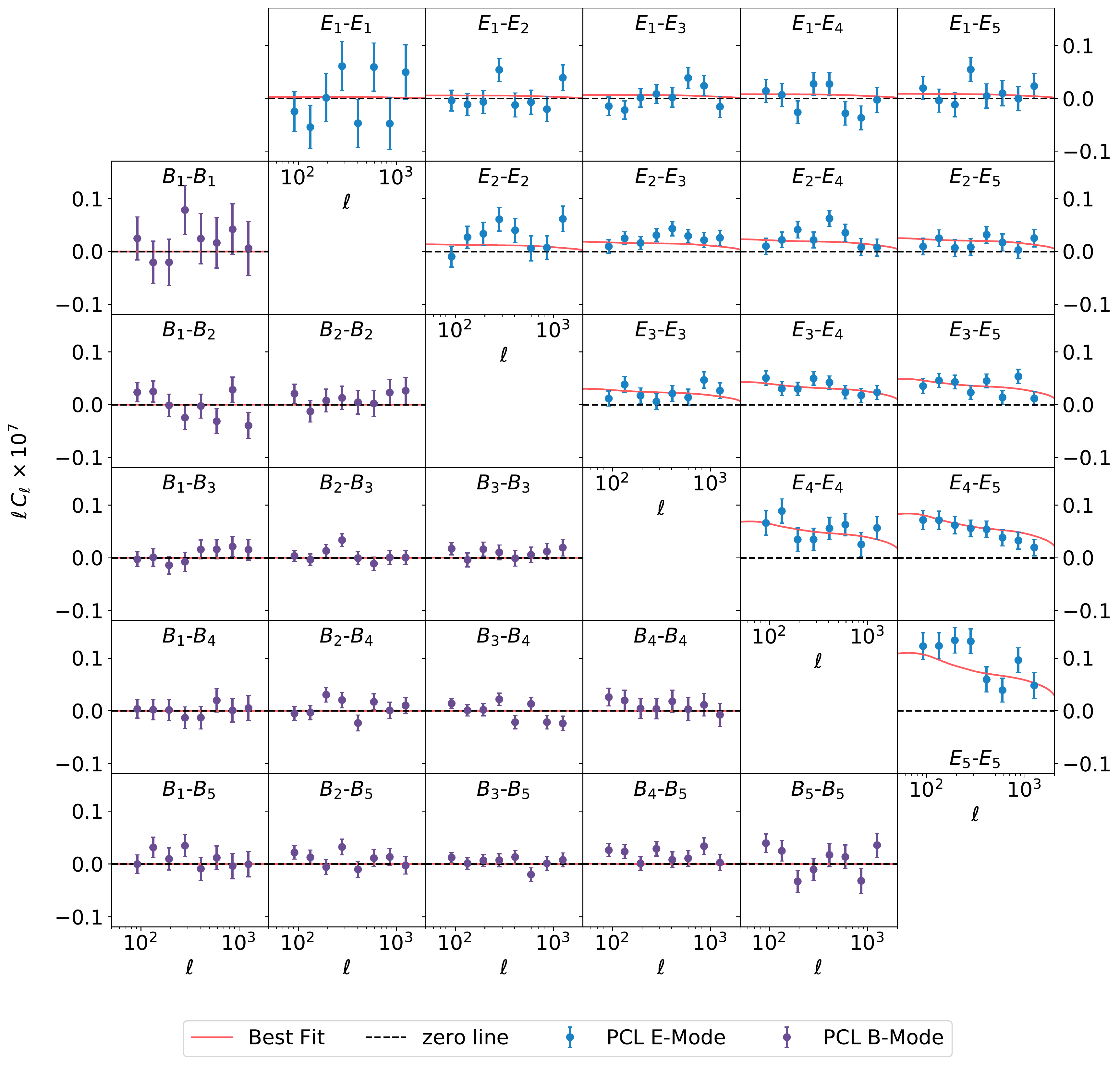}
      \caption{KiDS-1000 E-mode (upper triangle) and B-mode (lower triangle) \pcl{} measurements with the best-fit model from our cosmological analysis in Sect. \ref{Sec:Posteriors} convolved with the data's mixing matrix (Eq.~\ref{Eq:ConvCls}). The auto- and cross-angular power spectra have been measured in eight log-spaced bandwidths from $76 \leq \ell \leq 1500$. The error bars are estimated from the covariance (described in Sect. \ref{Sec:CovMat}), calculated with \flask{} \citep{2016-Flask} and \salmo{} \citep{2020-KiDS1000-Methods} simulations. We include the best-fit theory for the pseudo B-modes using Eq. \eqref{Eq:MixMatCls} (red line); the amplitude is too small to be distinguishable from the zero line with a reduced $\chi^2 \sim 1$ for a null detection.}
         \label{Fig:PCL-Measurements}
\end{figure*}

Our \pcl{} implementation uses pixelised shear maps for each redshift bin using a \hp{} grid \citep{2005-Healpix}. For the map generation, we use  $N_{\text{side}} = 1024$ ($N_{\text{pix}} = 12N_{\text{side}}^2$), measuring multipoles between $76 \leq \ell \leq 1500$ with the method described above. The lower bound we used in this work contains larger scales than the band-power measurements from \cite{2020-Asgari-2ptsK1000}\footnote{For comparison, the scales used in \cite{2020-Asgari-2ptsK1000} for band-power measurements were $100 \leq \ell \leq 1500$.}.
Observing the binned mixing matrices shown in Fig. \ref{Fig:Windows}, we note that by cutting at an bellow an $\ell_{\rm min} = 100$, we find that just a small number of lower multipoles get mixed into the scales of interest. Therefore, we decreased the $\ell$ lower bound towards larger scales. Although this choice is driven by survey geometry as below these scales practically no information is in the data, the specific value of $\ell_{\rm min} = 76$ was chosen arbitrarily. We were able to push to slightly lower $\ell$ values than the band-powers analysis from \citep{2020-Asgari-2ptsK1000} as the latter suffers strong mixing bellow $\ell=100$. Figure~\ref{Fig:Windows} also gives us an intuition for the $k$-scales that enter our analysis as a function of redshift and multipole given the bandwidth $\ell$-bins used.

Following \cite{2005-BrownCastroTaylor}\footnote{We note that this choice of weight could be optimised for cosmic shear analysis, which will be investigated in future work.}, the measured \pcl{}s are then binned using eight log-spaced multipole bins between $76 \leq \ell \leq 1500$,

\begin{align}
    \tilde{C}^{ij}_L = \frac{1}{2\pi}\sum_{\ell} \frac{\ell(\ell + 1)}{(\ell^{L + 1} - \ell^{L})} \tilde{C}_{\ell}^{ij}\, ,
    \label{Eq:BinnedCls}
\end{align}
where $\ell^{L} < \ell \leq \ell^{L + 1}$ are the boundaries of the bandwidth bins centred in $L$, shown in Fig. \ref{Fig:Windows} as the vertical dotted lines. The measured \pcl{} estimates are shown in Fig. \ref{Fig:PCL-Measurements} for the E- and B-modes, together with the best-fit theory line from the cosmological constraints shown in Sect. \ref{Sec:Posteriors}. The \pcl{} B-modes shown in the lower triangle part of Fig. \ref{Fig:PCL-Measurements} are consistent with zero with a $\chi_{\rm red}^2 = 1.07$ for a null-detection, indicating that even with the mode-mixing introduced by the mixing matrix, the contribution of E-modes mixed into B-modes is not enough to bias the cosmological measurements. The auto-power spectrum for $E_2$ and combinations with it, mostly $E_2-E_3$, have a slightly higher amplitude than the best-fit theory lines. A similar effect was found in previous KiDS-1000 studies but \citet{2020-Asgari-2ptsK1000} have shown that this has almost no impact on the cosmological constraints nor the goodness-of-fit, as shown in Fig.~7 of \citet{2020-Asgari-2ptsK1000}.

The number of multipole bins was arbitrarily selected so that the overlaps between the binned mixing matrices (filters), shown in Fig. \ref{Fig:Windows}, did not exceed 10\% of their integrated areas. This was done to prevent excessive correlations between adjacent multipole bins. However, it should be pointed out that since we estimate the covariance matrix used for our cosmological analysis using simulations (see Section \ref{Sec:CovMat} for details), we expect any residual correlations caused by these overlaps to be well described in our analysis.


\subsection{Simulations and covariance matrix estimation}\label{Sec:CovMat}

While analytical covariance models are widely used in cosmic shear \citep{2004-Efstathiou-PCL,2008-Joachimi-Covs, 2017-Cosmolike,2019-Alonso-naMaster,2019-Hikage-HSC, 2019-GarciaGaria-PCL-Cov, 2019-Li-Covariance, 2021-Krause-DES-Y3-Methods, 2020-Asgari-2ptsK1000, 2020-KiDS1000-Methods, 2020-Heymans-KiDS1000-Cosmology, 2020-Nicola-PCL} and it will be used for future \textit{Euclid} applications \citep{2021-Upham-PCLCov}, the forward-modelling approach for the mixing matrix makes it convenient to estimate the covariance matrix from simulations. We decided to use the \textit{Egretta} validated suite of 1000 log-normal simulations generated with \flask{}\footnote{\url{http://www.astro.iag.usp.br/~flask/}} \citep{2016-Flask} and \salmo{}\footnote{\url{https://github.com/Linc-tw/salmo}} presented in \cite{2020-KiDS1000-Methods}. This suite of mocks includes a number of known observational complexities such as variable depth source redshift distributions, and spatially varying galaxy sample properties. \flasmo{} mocks like \textit{Egretta} were used to validate the covariance matrix modelling in \cite{2020-KiDS1000-Methods}, as well as cosmic shear signal recovery, demonstrated in \cite{2020-Heydenreich} using semi-analytic models. Contrary to N-Body simulations, log-normal simulations are much faster to obtain, allowing for quick survey simulations that accurately reproduce the desired 2pt functions in just a few CPU hours. Figure~D.9 from \cite{2020-KiDS1000-Methods} shows the agreement between the covariances obtained by \textit{Egretta} mocks and the analytical covariances. Both methods agree to a few percent for cosmic shear 2pt functions, with the largest deviations coming from the first redshift bin. In other words, we are confident that our log-normal simulation-based covariance is accurate and that it reproduces, up to a few percent difference, what is expected by theoretical covariance calculations for cosmic shear analysis.

\begin{table*}
\caption{Comparison between the original mocks, the sub-sampled mocks and the Gold Sample SOM} 
\label{table:mocks}      
\centering          
\begin{tabular}{c c c c c c c}    
\hline\hline       
                      
Bin &  $\bar{z}$ (J21) & $\bar{z}$ (Sub-sampled) & $\bar{z}$ (Gold Sample) & $n_{\text{eff}}$ (J21) & $n_{\text{eff}}$ (Sub-sampled) & $n_{\text{eff}}$ (Gold Sample) \\
\hline                    
   1 & 0.39 &  0.40 & 0.26 & 0.85 & 0.62 & 0.62\\  
   2 & 0.49 &  0.47 & 0.40 & 1.56 & 1.19 & 1.18\\
   3 & 0.67 &  0.62 & 0.56 & 2.23 & 1.86 & 1.85\\
   4 & 0.83 &  0.80 & 0.79 & 1.52 & 1.27 & 1.26\\
   5 & 1.00 &  0.96 & 0.98 & 1.39 & 1.32 & 1.31\\
\hline
\end{tabular}
\tablefoot{Here we present a comparison between characteristics from the sub-sampled and original redshift distributions from \textit{Egretta} mocks with the Gold Sample SOM $N(z)$s from \citep{2020-Hildebrandt-SOM-KiDS1000}. The original mock values from \cite{2020-KiDS1000-Methods} are labelled as J21 and values labelled as \textit{sub-sampled} are the ones used in this work.}
\end{table*}

\textit{Egretta} mocks were generated using a combination of log-normal matter distribution simulations from \flask{} with post-processing using \salmo{} to populate the matter density field with galaxies based on their local redshift distribution and properties. The mock creation begins by generating the dark matter density field using angular power spectra calculated from 18 redshift tomographic bins of similar intervals, in comoving distances of 150 -- 200 Mpc$\, h^{-1}$, using the fiducial cosmology presented in Table A.1 from \cite{2020-KiDS1000-Methods}. \,\flask{} then integrates along the line-of-sight to obtain the shear and convergence fields. Using \salmo, galaxies are positioned in the sky according to the underlying matter density field and following a Poisson distribution, with their shapes given by the weak lensing fields generated by \flask. In addition, \salmo{} allows for spatially variable redshift distributions to be implemented, reproducing variable depth effects in the final simulated catalogues (more details in Section 4 of \citealt{2020-KiDS1000-Methods}).

Even though the \textit{Egretta} mocks were created with KiDS-1000 analysis in mind, they were originally generated using the KiDS+VIKING-450 (KV450) DIR redshift distributions \citep{2017-Hildebrandt-KiDS450,2020-Hidelbrandt-KV450}. In order to use these mocks for our purposes, we have sub-sampled them to match the KiDS-1000 SOM redshift distributions \citep{2020-Hildebrandt-SOM-KiDS1000}. The KV450 galaxy sample uses all the galaxies available, whereas in KiDS-1000 we remove galaxies using the SOM-based Gold sample. For this reason, we can sub-sample the \textit{Egretta} mocks; we have a lower $n_{\rm eff}$ for KiDS-1000 than for KV450. Table \ref{table:mocks} shows the sub-sampling from the original mocks to match the KiDS-1000 gold catalogue galaxy density. Although the mean redshift, $\bar{z}$, is slightly higher than the SOM's redshift distributions for the first two tomographic bins, the effective number of galaxies, $n_{\text{eff}}$, matches quite well. As higher mean redshifts lead to a higher variance in the sample's \pcl{}s, the differences in the mean redshifts yield more conservative errors  and, therefore, are not a concern.

The mocks were originally created at $N_{\text{side}} = 4096$; however, we apply the \pcl{} estimator at $N_{\text{side}} = 1024$ to match the process applied to the data (see Sect. \ref{Sec:PCL} for details), measuring angular power spectra using the same eight log-spaced $\ell$-bins. The difference in resolution is taken into account by correcting for the pixel window function at the target $N_{\rm side} = 1024$ \citep{2005-Healpix}. Lastly, we calculate the covariance of the noise subtracted pseudo-$C_{\ell}$s for all 1000 \textit{Egretta} mocks. It is fundamental that the noise estimates from each mock realisation are calculated in the same way as in the data: for each mock, we re-orient galaxies in the catalogue, calculate the noise angular power spectrum to obtain a noise realisation, and subtract the average of 100 noise realisations from the measured \pcl{} for that mock catalogue. We also note that the number of noise realisations used to estimate the noise for each of the mocks can have a significant impact on the covariance matrix, which directly impacts the cosmological constraints since we have a noisy estimate of the covariance. We performed tests using only 20 realisations and the impact on cosmological parameters was significant with an increase as high as 28\% on the $S_8$ error-bars. Therefore, we stress that a high number of noise power spectra estimates is crucial for accurate estimation of the covariance matrix.

We also verified that, for the mocks, the impact of the mixing matrix variation between each mock realisation is subdominant in the \pcl{}s. We are forward-modelling the mixing matrix into the theory, meaning that we include the partial sky effects into the mock's measurements. Although the mask varies slightly from mock to mock, the \pcl{}s effects due to the masking vary within the mock’s cosmic variance and noise. We calculated the variation of the mixing matrix from a sub-sample of the mocks and verified that, overall, they are similar to the data’s mixing matrix, with a maximum of a few percent difference. Moreover, we also verified that the variation is subdominant when convolving theory-vectors with the mocks’ mixing matrix and comparing with theory-vectors convolved with the data’s mixing matrix, showing a few percent difference.

The \textit{Egretta} mocks have no multiplicative bias, being equivalent to the KiDS-1000 data after the \textit{m}-correction is applied. The \textit{m}-correction has an uncertainty associated with it, which needs to be included in the covariance as an additional term. To propagate correctly the effects of the multiplicative shear calibration into our covariance matrix we take an analytical approach \citep{2020-KiDS1000-Methods}, modelling the \textit{m}-correction contribution to the covariance matrix as \citep{2002-Bridle-MCalibration, 2010-Taylor-MCorrection, 2021-Doux-MCorrection, 2020-KiDS1000-Methods}

\begin{align}
    {\rm Cov}_{m }\left( \tilde{C}^{i,j}_{\ell};\tilde{C}^{i',j'}_{\ell'} \right) = & \,\tilde{C}^{i,j}_{\ell}\tilde{C}^{i',j'}_{\ell'} \left[\sigma_{m }^{i}\sigma_{m }^{i'} + \sigma_{m }^{i}\sigma_{m }^{j'}\right. \nonumber \\ 
    &\left. \quad + \, \sigma_{m }^{j}\sigma_{m }^{i'} + \sigma_{m }^{j}\sigma_{m }^{j'}\right]\, ,
    \label{Eq:m-corrCov}
\end{align}
where $\sigma_{m }^i$ is defined as the standard deviation of the \textit{m}-corrections shown in Table \ref{table:1} and $\tilde{C}^{E_iE_j}_{\ell}$ are theory angular power spectra convolved with the mixing matrix using Eq. \eqref{Eq:ConvCls}. This expression assumes that the expectation value of \textit{m} is zero after the correction, with $\sigma^i_{m } \ll 1$. Since the $\sigma_{m }$ part of Eq. \eqref{Eq:m-corrCov} is independent of $\ell$, the mixing of modes does not affect this part of the covariance. \cite{2020-Asgari-2ptsK1000} demonstrated that this approach gives equivalent results to having the $m_{ z}$ correction as a free parameter, for each redshift tomographic bin \textit{z}, during the cosmological inference analysis.

\begin{figure}
   \centering
   \includegraphics[width=\columnwidth]{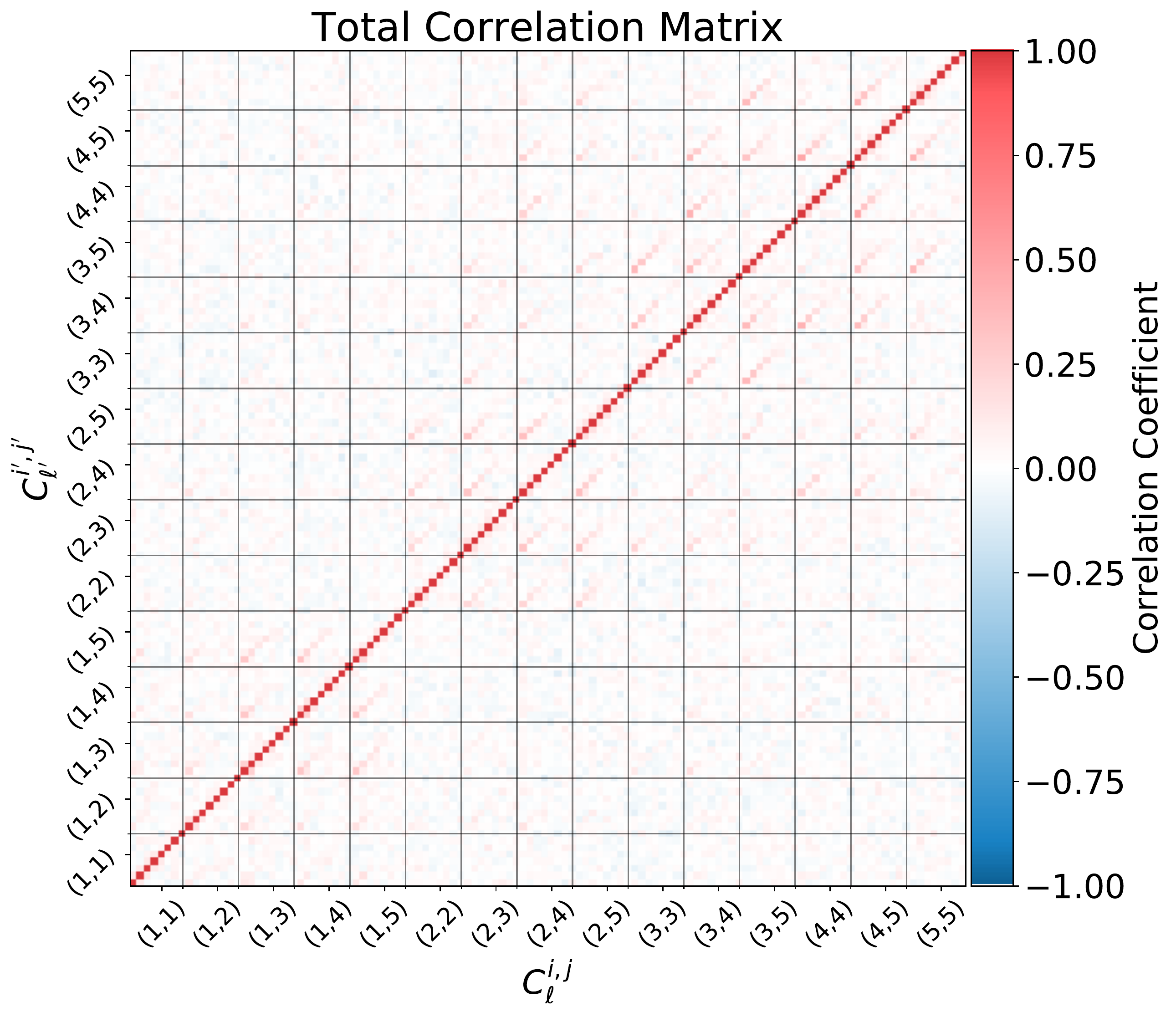}
      \caption{Correlation matrix for the E-mode \pcl{} covariance. The covariance matrix was calculated using 1000 mocks from the \textit{Egretta} suite of \flasmo{} simulations \citep{2020-KiDS1000-Methods}. The indices $(i,j)$ in the labels are the redshift bin numbers for each pair of $C^{i,j}_{\ell}$ E-modes.}
         \label{Fig:Cov}
\end{figure}

Finally, our total covariance matrix is the combination of the simulated covariance from the \textit{Egretta} mocks and the analytic m-correction covariance: $\rm{Cov}_{\rm tot}  = \rm{Cov}_{\rm mocks} + {\rm Cov}_{\textit{m}}$. The resulting covariance, $\rm{Cov}_{\rm tot}$, is presented as a correlation matrix in Fig. \ref{Fig:Cov} and shows a similar structure as previously found by \cite{2020-KiDS1000-Methods} for the band-powers analysis.

\section{Systematic contamination analysis}\label{Sec:Systematics}
\begin{figure}
  \centering
  \includegraphics[width=\columnwidth]{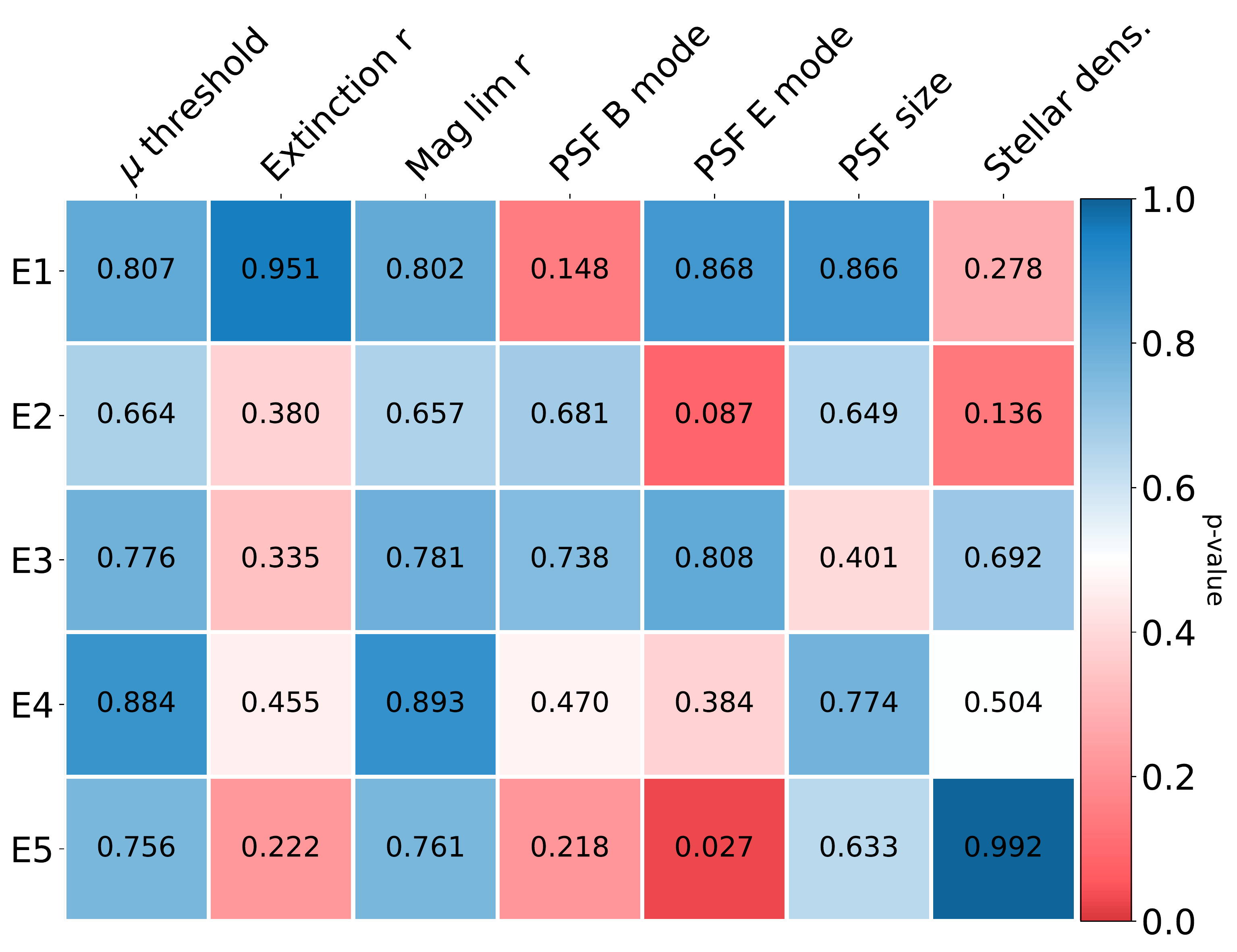}
      \caption{P-values for the cross-correlations between the data's tomographic pseudo E-modes and different systematics for a null detection ($p \leq 0.05$ would indicate a detection). Here we show the systematics motivated in Sect. \ref{Sec:Systematics}: object detection threshold ($\mu$-threshold), extinction for the \textit{r}-band, magnitude limit for the \textit{r}-band, PSF ellipticities components in harmonic space (E/B-mode), the PSF size, and stellar overdensity from Gaia DR2 \citep{2018-Gaia}. The highest correlation occurs for the PSF E-mode in the fifth tomographic bin, this is shown in Fig. \ref{Fig:Syst-PSF-E-E5} to be subdominant to the cosmological signal given the estimated statistical error in this particular tomographic bin.}
         \label{Fig:Syst-RedChi2-E}
\end{figure}

Photometric galaxy surveys are affected by observational effects related to instrumental, astrophysical, and atmospheric phenomena. When measuring the shapes and positions of galaxies, calibration and control of such systematic effects are at the core of a successful analysis. Like any galaxy survey, the KiDS-1000 data set experiences systematic contaminations given the complex nature of combining optical and infrared data spread through nine different bands. For example, instrumental effects such as the telescope's point spread function can vary due to telescope inclination, impacting galaxy shape measurements resulting in the introduction of spurious and artificial correlations in the final data-vector. Other observational effects like a spatially varying magnitude limit in a given band also have the potential to contaminate or bias the sample, inducing an artificial selection, for example. Finally, astrophysical phenomena like Galactic extinction and stellar contamination in the galaxy sample can impact the quality of data, resulting in correlations that alter the data's measured angular power spectra leading to a potential bias in the inferred cosmology.

An effort to select, clean, and validate this sample was undertaken by \cite{2020-KiDS-1000-ShearCat}, including a range of contamination checks. In this section, our goal is to analyse any possible residual systematic contamination in the KiDS-1000 catalogue that could affect the measured angular power spectra of galaxy ellipticities. We perform null tests by cross-correlating data with relevant observational quantities in the catalogue and stellar density from Gaia Data Release 2 \citep{2018-Gaia}. We then observe the goodness-of-fit to a zero-correlation case. The approach taken here is similar to null tests in \cite{2014-Leistedt-Systematics}, but without the use of mode-projection.

A summary of observational, astrophysical, and instrumental systematics analysed is as follows:
\begin{itemize}
    \item \textbf{Extinction ($ugriZYJHK_s$)}: Galactic extinction for each of the observed nine bands derived from the $E(B-V)$ maps from \cite{1998-Schlegel-Extinction} combined with the extinction coefficients, $R_V = 3.1$, from \cite{2011-Schlafly}. The extinction coefficients for each filter, $R_f$, follow a linear relation between the observed bands: $R_f = A_f/E(B-V)$. Therefore, we have only shown results for the \textit{r}-band since this is the detection band for KiDS-1000 and other values are just linearly scaled based in this band. The values for $R_f$ used for KiDS-1000 can be found in Table~4 of \cite{2019-KiDS-DR4}.\\[0.01cm]
    \item \textbf{Magnitude limit ($ugriZYJHK_s$)}: magnitude limit in each of the observed nine bands. This quantity follows a forced photometry based on the \textit{r}-band detection. We only show results for the \textit{r}-band as it is the detection band used for forced photometry. \\[0.01cm]
    \item \textbf{$\mu$-Threshold}: object detection threshold above the background.\\[0.01cm]
    \item \textbf{PSF ellipticities}: mean PSF ellipticity components, treated as a spin-2 field, converted to \textbf{PSF E/B-modes} in harmonic space (see Eqs. \ref{Eq:PseudoE} and  \ref{Eq:PseudoB}). The value quoted in Fig. \ref{Fig:Syst-RedChi2-E} has the PSF $\alpha$-correction term from \cite{2020-KiDS-1000-ShearCat} applied to the PSF ellipticities before converting them to harmonic space. The subsequent text explores and explains this in detail.\\[0.01cm]
    \item \textbf{PSF Size}: The trace of the PSF magnitude moments, $T = Q_{11} + Q_{22}$, where $Q_{ij}$ is defined in terms of the second-order moments of the two-dimensional angular light distribution, $I(\hat{\mathbf{n}})$; see Sect.~3.1 of \citet{2020-KiDS-1000-ShearCat}. \\[0.01cm]
    \item \textbf{Stellar Density}: star sample from Gaia DR2\footnote{Obtained from \url{https://cds.u-strasbg.fr/gaia}} containing all Gaia DR2 objects in a \hp{} grid of $N_{\rm side} = 1024$.
\end{itemize}

For each of the quantities mentioned above, we create \hp{} maps using the catalogue object positions with $N_{\rm side} = 1024$ and the mean value for the observables in each pixel. A spherical harmonic transform is applied to the maps --- spin-0 for all scalar quantities and spin-2 transforms for the PSF ellipticities ---  in order to obtain the systematic's spherical harmonic coefficients, $a_{\ell m}^{\text{syst}}$, and the data's spherical harmonic coefficients, $X_{\ell m}^{i}$ -- where $\text{X} = \{E, \, B\}$ and \textit{i} is the tomographic bin index. Using the spherical harmonic coefficients, we estimate the cross-power spectra between data and the systematic quantities of interest as

\begin{equation}
        \tilde{C}^{\text{X}_i,\text{syst}}_{\ell} = \frac{1}{(2\ell +1)} \sum_m \text{X}_{\ell m}^{i} \left(a_{\ell m}^{\text{syst}}\right)^*\, .
        \label{Eq:Cls_Syst}
\end{equation}

To estimate the errors for the cross-correlations between data and systematics, we obtain covariances in two different ways. For the catalogue quantities, we use a jackknife re-sampling method. We start by subdividing the survey footprint into 60 distinct regions using a K-Means algorithm on the sphere \citep{2017-DES-SV-KMeans}\footnote{Our K-Means code uses the code \texttt{kmeans\_radec}, developed by Erin Sheldon which can be found here: \url{https://github.com/esheldon/kmeans_radec}.}. While removing the \textit{k}th region, we calculate the \pcl{} (Eq. \ref{Eq:Cls_Syst}) of each sub-sample for all 60 regions in the footprint. For the stellar density Gaia map, we take a different approach. The KiDS-1000 footprint is naturally over a low stellar density region in the sky, meaning that the jackknife covariance method leads to very small variations in the cross-power spectra. This is due to most of the jackknife regions containing a very small number of stellar objects in the Gaia DR2 map, if any. {Therefore, the covariances for the cross-power spectra between stellar density and the data are calculated using the cross-correlations between the Gaia DR2 map and 200 noise realisations of the galaxy ellipticities in the catalogue (created by randomly orienting the galaxies in the KiDS-1000 catalogue for each realisation). Given the stochastic nature of this process, it yields a somewhat underestimated covariance, which leads to a conservative null test.}

\begin{figure}
  \centering
  \includegraphics[width=\columnwidth]{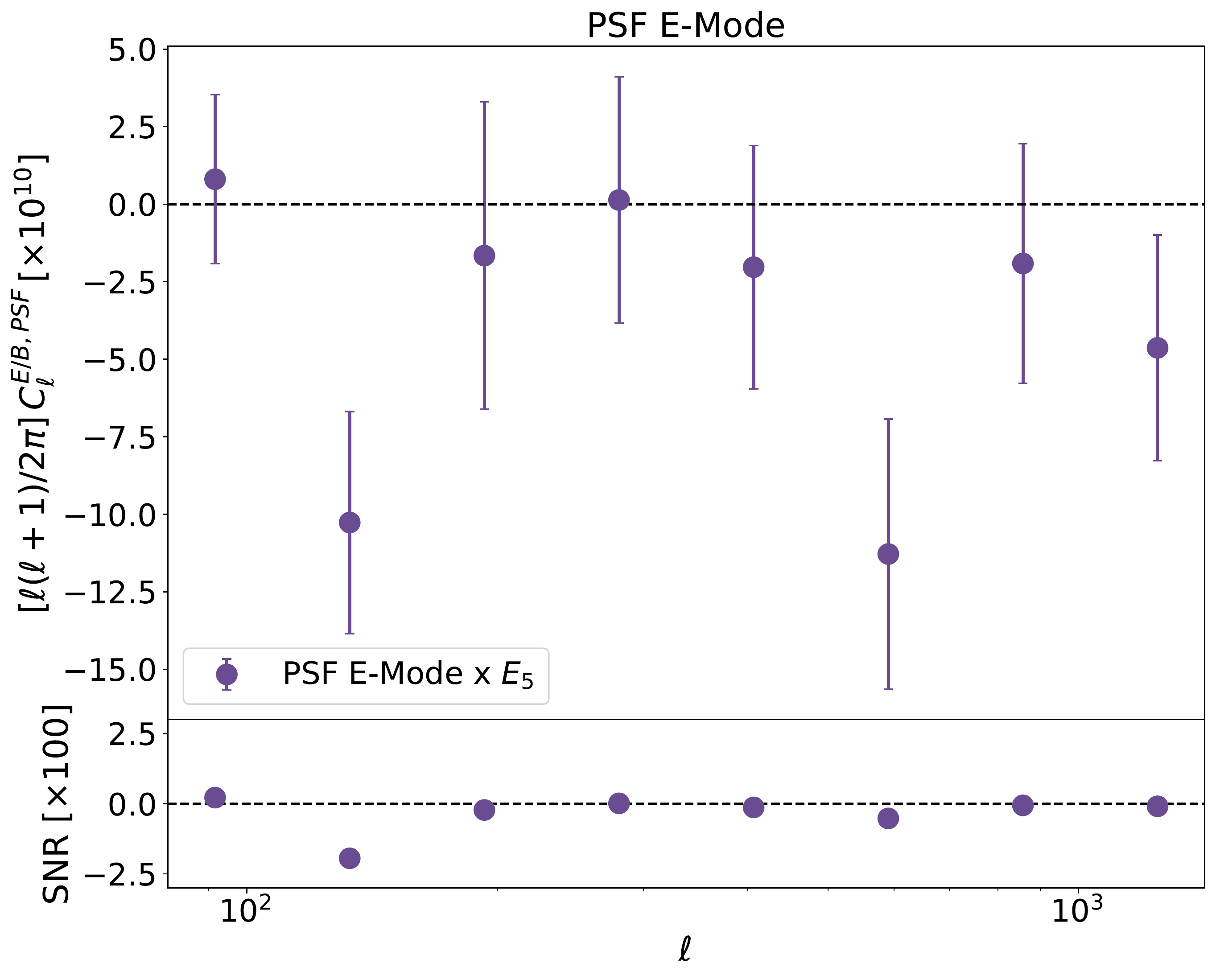}
      \caption{PSF E-mode and $E_5$ cross-power spectrum. {(Top panel)} Cross-angular power spectra between PSF E-mode component and $E_5$ with the $\alpha$-correction from \cite{2020-KiDS-1000-ShearCat} applied to the PSF ellipticities. {(Bottom panel)} Signal-to-noise ratio as defined by Eq. \eqref{Eq:SNR}, showing that the largest S/N multipole bandwidth has $\text{S/N} \sim -2.5\times 10^{-2}$, meaning that it is well within the estimated covariance for our analysis.}
         \label{Fig:Syst-PSF-E-E5}
\end{figure}

Subsequently, we calculate the goodness-of-fit using the p-value for a fit to zero correlations between systematics and measured ellipticities. This is shown for the E-modes in Fig. \ref{Fig:Syst-RedChi2-E} and in Fig.  \ref{Fig:Syst-RedChi2-B} for the B-modes as heat-maps. From Fig. \ref{Fig:PCL-Measurements} we do not detect any \pcl{} B-modes; therefore, the B-modes systematics analysis are shown in Appendix \ref{Apdx:BModesSyst}. 

Initially, we find a significant correlation with the PSF ellipticity and the 5th tomographic E-mode, with  a
p-value of $0.004$ for a cross-correlation detection. This result is to be expected from the analysis of \cite{2020-KiDS-1000-ShearCat}, who quantified the level of contamination to the cosmic shear signal from PSF leakage. \cite{2020-KiDS-1000-ShearCat} measured the $\alpha$ parameter per tomographic bin, where $\alpha$ quantifies the average fraction of PSF ellipticity in the shear estimator. As the KiDS PSF has a very low ellipticity on average, and as $\alpha$ was found to be small (at the level of 4\% for $z_{\rm B} > 0.7$), or consistent with zero (for $z_{\rm B} < 0.7$), \cite{2020-KiDS-1000-ShearCat} concluded that the statistically significant systematic PSF leakage introduced less than a $0.1\sigma$ in the inferred $S_8$ constraints for KiDS-1000, inducing no significant biases.

In this analysis, we recover the angular dependency on this systematic using the $\alpha$ correction for the PSF ellipticity measurements, finding a cross-correlation between the fifth tomographic bin E-mode with a p-value of  $0.027$ for a non-zero signal. Although the value is not worryingly {low}, the fifth tomographic bin contains most of our constraining power \citep{2020-Asgari-2ptsK1000,2020-Heymans-KiDS1000-Cosmology}. Therefore, it is crucial to understand if such correlations could indeed mimic a cosmological signal and bias our final cosmological analysis. We show this cross-angular power spectrum in detail in Fig. \ref{Fig:Syst-PSF-E-E5} together with its signal-to-noise ratio (S/N, Eq. \ref{Eq:SNR}) for the cross-correlations between data and PSF E-mode and the data's covariance estimated in Sect. \ref{Sec:CovMat}. The S/N demonstrates that this cross-correlation is subdominant in the E-mode signal from the fifth tomographic bin, with the bandwidth with the highest S/N around $5\times10^{-2}$. This means that this excess correlation is subdominant given the data's estimated covariance in this tomographic bin, leading us to draw the same conclusions as \cite{2020-KiDS-1000-ShearCat} in that this correlation has a sufficiently low significance that it has no impact on the cosmological parameter inference.

In summary, our conclusion from the examined potential systematics is that no considerable contamination is biasing the measured E-mode angular power spectra shown in Fig. \ref{Fig:PCL-Measurements}. Therefore, the estimated cosmological parameters shown in Section \ref{Sec:Posteriors} are purely a result of the cosmological and astrophysical signals captured by our measured \pcl{}s.

\section{Cosmological inference}\label{Sec:Inference}
This section describes the final elements necessary for cosmological inference using tomographic cosmic shear angular power spectra data: theory forward modelling, the likelihood and the priors we use to obtain the posteriors presented in Section \ref{Sec:Posteriors}. Here, we also describe the external clustering data included to obtain combined constraints using galaxy clustering information from BOSS \& eBOSS with Lyman-$\alpha$ forest data from eBOSS.

Cosmological inference is performed using a version of the \texttt{Monte Python 3} software \citep{Monte-Python-Audren:2012wb,Monte-Python-Brinckmann:2018cvx}\footnote{\url{https://github.com/brinckmann/montepython_public}} modified to directly sample from Gaussian priors\footnote{The modified version sampling from Gaussian priors can be found at: \url{https://github.com/BStoelzner/montepython_public/tree/gaussian_prior}} for the \texttt{Multinest} sampler \citep{2008-Multinest1, 2009-Multinest2, 2014-pyMultinest, 2019-Multinest3}. It has also been modified to sample directly from the $S_8$ parameter according to the findings reported in \cite{2020-KiDS1000-Methods}. The \texttt{Monte Python 3} pipeline was compared with \texttt{KCAP}, the pipeline used in \cite{2020-Asgari-2ptsK1000} and \cite{2020-Heymans-KiDS1000-Cosmology}, and results were found to be consistent.

Following \cite{2020-KiDS1000-Methods}, we report our constraints in two different ways: the traditional marginalised constraints, extracted from  the 1D marginalised posteriors for each individual parameter, denoted marginal; and the {maximum a posteriori} (MAP) with the projected joint highest posterior density (PJ-HPD); this is also referred to as MAP+PJ-HPD. This method ensures, by definition, that the multi-dimensional MAP is always within the PJ-HPD region and it has been demonstrated to accurately report the posterior distributions for each parameter, impacting significantly the reported results from skewed distributions. \texttt{Monte Python} originally does not deal with Gaussian priors and its minimisation algorithm acts directly on the likelihoods, making the MAP estimates from minimising the log-posterior unreliable. We take a different approach than the one proposed by \cite{2020-KiDS1000-Methods} by finely sampling the posterior using a high number of live-points and obtaining the MAP estimates from the \texttt{MultiNest} chains instead of minimising the log-posterior. Tests show that the MAP values we recover have a better goodness-of-fit from this approach than the minimisation approach. This means our estimate of the MAP is not as accurate as it could be; however, there is no impact for the PJ-HPD boundaries.  For more details, see Section 6.4 from \cite{2020-KiDS1000-Methods}.

\subsection{Theoretical modelling}\label{Sec:Theory}
The modelling approach for the cosmic shear angular power spectra is very similar to the one described in \cite{2020-KiDS1000-Methods} with a few technical differences and additional steps related to the forward-modelling of the mixing matrix. Differently from \cite{2020-KiDS1000-Methods}, \texttt{Monte Python 3} obtains the underlying matter power spectra from \texttt{Class} \citep{2011-CLASS-1, 2011-CLASS-2}.\footnote{\url{https://lesgourg.github.io/class_public/class.html}} Following the other KiDS-1000 analyses, we fix the sum of neutrino masses to $\sum m_{\nu} = 0.06 {\rm eV}c^{-2}$, using the one massive species approximation of the normal hierarchy. Not only was this the approach taken by \cite{Planck2018-cosmology}, but it was shown in \cite{2019-LoureiroNeutrinos} that it has no impact on the standard flat \lcdm{} cosmological parameters compared to more complex neutrino mass models.

The non-linear matter power spectrum, $P_{\rm m}^{\rm nl}(k)$, is calculated using the halo model from \texttt{HMCode} \citep{2015-Mead-HMCode} included in \texttt{Class}.\footnote{\cite{2021-HMCode} introduce a new version of the \texttt{HMCode}, with improved accuracy. \cite{2020-Tilman-K1000-Ext} compared this new \texttt{HMCode} against the version used in this work and found that it introduces a $0.26\sigma$ shift in $S_8$. This exemplifies how the non-linear modelling is the limiting theoretical systematic in cosmic shear analysis.} This approach incorporates AGN baryonic feedback into the matter power spectrum using a model with two parameters: the halo bloating parameter, $\eta_0$, and the halo mass-concentration relation amplitude, $A_{\rm bary}$. These parameters can be  constrained by the following relation from \cite{2018-Joudaki}: $\eta_0 = 0.98 - 0.12 A_{\rm bary}$. Therefore, $A_{\rm bary}$ can be used as a single free parameter to model baryonic effects. 

Subsequently, the tomographic cosmic shear angular power spectra are modelled using projections along the line-of-sight of the 3D matter power spectrum in redshift bins \textit{i} and \textit{j}. The observed $C_{\ell}$s are a combination of contributions from cosmic shear, indexed by $\gamma$, and the intrinsic alignment of galaxies, indexed by $\text{I}$ \citep{2009-Bernstein, 2010-JoachimiBridle}:

\begin{align}
\label{Eq:ClsCombination}
    C^{\epsilon_i, \epsilon_j}_{\ell} = C^{\gamma_i, \gamma_j}_{\ell} + C^{\gamma_i, \text{I}_j}_{\ell} + C^{\text{I}_i, \gamma_j}_{\ell} + C^{\text{I}_i, \text{I}_j}_{\ell}\, .
\end{align}
Using the Limber approximation \citep{1992-Kaiser, 2008-Loverde, 2017-Kitchin-janemseimaispq, 2017-Kilbinger-LimitsCS,2017-Lemos-Limber}, where the wavenumber \textit{k} and the radial co-moving distance $\chi$ can be related as $kf_{\rm K}(\chi) = \ell + 1/2$, we can express the right-hand side $C_{\ell}$s from Eq. \eqref{Eq:ClsCombination} as 
\begin{align}
    C_{\ell}^{X_i, Y_j} = \int_0^{\chi_{\rm H}}{\rm d}\chi \frac{W_X^i(\chi)W_Y^j(\chi)}{f_{\rm K}^2(\chi)}P_{\rm m}^{\rm nl} \left( \frac{\ell+1/2}{f_{\rm K}(\chi)}, z(\chi) \right)\, ,
    \label{Eq:ClsTheory}
\end{align}
where the $X, Y$ indices are $\gamma$ and/or I, the $i,j$ indices are the tomographic bins indices for the different kernels, $\chi_{\rm H}$ is the co-moving distance to the horizon and $f_{\rm K}(\chi)$ is the Friedmann-Lema\^{i}tre-Robertson-Walker (FLRW) co-moving angular diameter distance. Note, also, that we only consider flat \lcdm{} models in this study, therefore, $f_{\rm K}(\chi) = \chi$. The weak lensing kernel is defined as:
\begin{align}
    W_{\gamma}^i (\chi) = \frac{3}{2} \frac{H_0^2\Omega_{\rm m}}{c^2a(\chi)}f_{\rm K}(\chi)\int_{\chi}^{\chi_{\rm H}}{\rm d}\chi'n^i_s(\chi')\frac{f_{\rm K}(\chi' - \chi)}{f_{\rm K}(\chi')}\, .
\end{align}

The intrinsic alignments kernel is calculated using the  non-linear alignment (NLA) model from \cite{2007-BridleKing} without redshift evolution,

\begin{align}
    W^i_{\rm I}(\chi) = -A_{\rm IA} \frac{C_1\rho_{\rm cr}\Omega_{\rm m}}{D[a(\chi)]}n_s^i(\chi)\, ,
\end{align}
with $C_1\rho_{\rm cr}\approx 0.0134$ \citep{2011-Joachimi} and where $D[a(\chi)]$ is the growth function and $A_{\rm IA}$, the amplitude of intrinsic alignments, kept as a free parameter in the analysis but the same for all redshift tomographic bins. Although not physically motivated in the way non-linearities are accounted for, the NLA model was found by \cite{2021-Fortuna} to be precise enough for the case we are currently analysing.

The theory angular power spectra from Eq. \eqref{Eq:ClsCombination} are then corrected for the pixel window function \citep{2005-Healpix}, convolved with the mixing matrix using Eq. \eqref{Eq:ConvCls}, assuming $C_{\ell}^{B_i,B_j} = 0$, and binned in eight log-space bandwidths between $76 \leq \ell \leq 1500$ using Eq. \eqref{Eq:BinnedCls}. Although not individually binned for the analysis, the binned mixing matrix can be seen as a band-pass filter, showing which scales get mixed for each of the log-spaced $\ell$-bins used in the data as in Fig. \ref{Fig:Windows}. This final convolved and binned theory vector is then compared to the measurements in the likelihood for cosmological inference.


\subsection{External data from SDSS}
\label{Sec:ExternalData}
Since our main objective in this study is to extract cosmological information from late-time probes using galaxy surveys, we have combined our cosmic shear measurements with clustering information. For the current generation of surveys, weak lensing alone has little constraining power beyond the $S_8$ parameter and it is limited by the degeneracy in the $\sigma_8\,-\Omega_{\rm m}$ plane \citep{2021-Hall-S8}. In order to compete with the extremely precise cosmic microwave background measurements from \textit{Planck}, this synergy between late-time probes is crucial. By combining late-time probes we can have a better understanding of the growth of structure tension with early-time probes. 

We use the publicly available SDSS likelihoods\footnote{\url{https://svn.sdss.org/public/data/eboss/DR16cosmo/tags/v1_0_0/likelihoods/BAO-plus}}, which combine pre-reconstruction full-shape\footnote{In this context, the so-called full shape means that the relevant BAO and RSD quantities are fitted to the full shape of the correlation function or power spectrum and then used to constrain cosmology.} results and post-reconstruction BAO results, both from the BOSS and eBOSS LRG samples \citep{2017-Alam,2020-Gil-Marin,2021-Bautista}. We also include BAO results from the Lyman-$\alpha$ (Ly$\alpha$) forest auto-correlation and its cross-correlation with quasar positions \citep{2020-Bourboux}. We chose to leave the quasar auto-correlations out, as sampling their biases could have an impact on the growth of structure parameter, $S_8$, since this parameter is degenerate with the bias.

Using \flasmo{} mocks, \cite{2020-KiDS1000-Methods} showed that the BOSS and KiDS-1000 data sets can be considered independent, with their cross-covariance consistent with zero. At the same time, even though the BOSS and KiDS-1000 footprints have a small overlap, \cite{2020-Heymans-KiDS1000-Cosmology} found that the galaxy galaxy-lensing correlation between the two has very little impact on the cosmological results, mostly due to the necessarily aggressive scale cuts in the data-vector due to bias and IA model limitations. Yet, it does have a small impact on the nuisance parameters, $A_{\rm bary}$ and $A_{\rm IA}$, as shown in Fig. 6 from \cite{2020-Heymans-KiDS1000-Cosmology}. Therefore, we also consider the two data sets, KiDS-1000 and BOSS, as independent and do not consider any galaxy-galaxy lensing correlations in this work. We use the already-existing \texttt{Monte Python} BOSS full-shape likelihood, but update it with eBOSS data. Following \cite{2020-eBOSSDR16}, we use BOSS results from the two low-redshift ($0.2<z<0.6$) LRG bins, and the higher redshift ($0.6<z<1$) eBOSS LRG bin that combines both BOSS and eBOSS data. For each of these bins we use measurements of $D_{\rm M}(z_\text{eff})/r_{\rm d}$, $D_{\rm H}(z_\text{eff})/r_{\rm d}$ and the rate of structure growth, $f(z_\text{eff})\sigma_8(z_\text{eff})$. Here, $D_{\rm M}(z)=(1+z)^{-1}f_{
\rm K}\left[\chi(z)\right]$ is the comoving angular diameter distance, $D_{\rm H}(z)=c/H(z)$, $r_{\rm d}$ is the size of the sound horizon at the end of the drag epoch, $f(z_\text{eff})$ is the linear growth rate, and $z_\text{eff}$ is the effective redshift of the measurement (as defined by Eq. 7, \citealt{2014-Ross-SDSS}). 

For the Ly$\alpha$ forest measurements we use the \texttt{Monte Python} DR14 likelihoods \citep{2019-Cuceu-BAOBBN}, but updated with DR16 data \citep{2020-Bourboux}. In this case we have measurements of $D_{\rm M}(z_\text{eff})/r_{\rm d}$ and $D_{\rm H}(z_\text{eff})/r_{\rm d}$, for both the auto-correlation and cross-correlation. We treat the two as independent, following a study on synthetic data sets \citep{2017-Bourboux} that found negligible correlations. From now on, we denote this combined data set as \textit{SDSS BAO and RSD}. 

Finally, we note that in contrast to the approach for the BOSS data used in \cite{2020-Heymans-KiDS1000-Cosmology} and \cite{2020-Tilman-BOSS-LCDM}, the SDSS likelihoods and data we use here do not assume a flat \lcdm{} model, that is they are not constraining cosmology using the full galaxy power spectrum. Instead, the SDSS likelihoods we are using in this work measure the quantities mentioned above -- $D_{\rm M}(z_\text{eff})/r_{\rm d}$, $D_{\rm H}(z_\text{eff})/r_{\rm d}$, $f(z_\text{eff})\sigma_8(z_\text{eff})$ -- fitted from the measured correlation function or power spectrum estimates, to constrain cosmology. This approach is considered to be `model-independent' as these quantities do not assume a flat \lcdm{} power spectrum when measured. Nonetheless, this `model-independent' approach sacrifices precision as pointed out by \cite{2020-Tilman-BOSS-LCDM}.

\subsection{Likelihood and priors}\label{Sec:LikePriors}

\begin{table}
\caption{Priors on cosmological and nuisance parameters.} 
\label{table:priors}      
\centering          
\begin{tabular}{l l l}    
\hline\hline       
Parameter name      &  Symbol & Prior                   
 \\
\hline                    
  Density fluctuation amp. & $S_8$ & [0.1, 1.3] \\
  Cold dark matter density & $\Omega_{\text{c}}h^2$ & [0.051, 0.255]\\
  Baryonic matter density & $\Omega_{\rm b} h^2$ & [0.019, 0.026] \\
  Hubble constant & $h$ & [0.64, 0.82]\\
  Spectral index & $n_{\rm s}$ & [0.84, 1.1]\\
  \hline
  Intrinsic alignment amp. & $A_{\text{IA}}$ & [$-6,6$]\\
  Baryonic feedback amp. & $A_{\rm bary}$ & [2,3.13]\\
  Redshift displacement & $\delta_z$ & $\mathcal{N}(\bm\mu, \tens{C}_{z})$\\
\hline
\end{tabular}
\tablefoot{The $\Omega_{\rm b} h^2$ prior, motivated by BBN constraints from \cite{2014-Olive-BBN}, is a conservative (5$\sigma$) equivalent to combining BBN data when external data from SDSS is used, similar to what is shown in \cite{2019-Cuceu-BAOBBN}. Here, $\mathcal{N}(\bm\mu, \tens{C}_{z})$ is a normal distribution with mean $\bm \mu$ and covariance $\tens{C}_{z}$ as estimated by \cite{2020-Hildebrandt-SOM-KiDS1000}.}
\end{table}  

In this section we discuss the likelihood and prior choices made for our analysis. Consistency with the main KiDS-1000 analysis \citep{2020-Asgari-2ptsK1000, 2020-Heymans-KiDS1000-Cosmology} is at the core of our choices. One of our objectives in this work is to show the accuracy of a pseudo-$C_{\ell}$ methodology with forward-modelling which could be applied to future Stage-IV galaxy surveys. We want to extend the comparison with other 2pt statistics estimators from \cite{2020-Asgari-2ptsK1000}. This means that making similar prior choices is important as it allows for an easy and fair comparison. 

Our covariance matrix is partially estimated from simulations with an analytical part for the \textit{m}-correction error propagation. Therefore, we opt for a Gaussian likelihood instead of a multivariate t-distribution \citep{Student-t-1931} as suggested by \cite{2016-SellentinHeavens} or the exact \pcl{} likelihood, presented in \cite{2020Upham-exact}, as this approach is more computationally expensive. This is motivated by the fact that with a sufficiently high number of simulations and for sufficiently high multipoles, the Gaussian approximation is a satisfactory representation of the underlying distribution for the cosmological parameters of interest \citep{2019-Taylor-LFI, 2020-Lin-NonGaussian, 2020Upham-Gaussian}.

The priors on all cosmological and nuisance parameters, with the exception of the redshift displacement parameters, are chosen to be flat and are detailed in Table \ref{table:priors}. The following paragraphs motivate and explain all prior choices. For more details, please refer to Appendix B from \cite{2020-Heymans-KiDS1000-Cosmology} and Sect. 6.1 from \cite{2020-KiDS1000-Methods}. 

As cosmic shear data are directly sensitive to the amplitude of density fluctuations, $\sigma_8$, or the growth of structure, $S_8$, this prior choice is of fundamental importance in our analysis \citep{2021-Hall-S8}. In \cite{2020-KiDS1000-Methods}, it was shown that probing this parameter as a derived parameter from the primordial power spectrum amplitude, $A_{\rm s}$ or $\ln (10^{10}A_{\rm s})$, can lead to small biases as the prior becomes informative for $S_8$. Therefore, in our analysis, we sample $S_8$ directly instead of treating it as a derived parameter. This guarantees that our prior on this parameter is flat.

Other parameter priors that are not properly constrained by cosmic shear data are chosen to be $\pm 5\sigma$ from independent measurements such as the prior on $h$, taken from the \cite{2016-RiessHubble} measurements, and the prior on $\omega_{\rm c} = \Omega_{\rm c} h^2$, derived from Supernova Type Ia constraints on $\Omega_{\rm m}$ from \cite{2018-Scolnic-SNa}. Nevertheless, there is one prior choice that has a significant impact when combining cosmic shear data with external galaxy clustering data from SDSS (see Sect. \ref{Sec:ExternalData} for more details): the baryonic matter density, $\omega_{\rm b} = \Omega_{\rm b} h^2$. This prior is chosen to reflect $\pm 5\sigma$ constraints from big bang nucleosynthesis (BBN) presented by \cite{2014-Olive-BBN}. This is a conservative choice for cosmic shear-only analysis as justified by \cite{2020-Heymans-KiDS1000-Cosmology} and \cite{2020-Asgari-2ptsK1000}. When combining cosmic shear data with clustering information from BOSS/eBOSS galaxy clustering and the eBOSS DR16 Ly$\alpha$ forest, this prior is a conservative equivalent to adding BBN data to our analysis, as shown in \cite{2019-Cuceu-BAOBBN}. Therefore, we would like to stress that the combined constraints between KiDS-1000 weak lensing and SDSS BAO and RSD shown in Sect. \ref{Sec:Posteriors} should be interpreted with the BBN-`inspired' priors in mind.

Finally, the redshift distributions for each of the five tomographic bins are allowed to shift according to a correlated Gaussian prior with $\bm\mu = (0.0001, 0.0021, 0.0129, 0.0110, -0.0060)$ and a covariance $\tens{C}_{z}$ calibrated in \cite{2020-Hildebrandt-SOM-KiDS1000} using mock galaxy surveys, according to the method presented by \cite{2020-Wright-SOM}. Constraints for the nuisance parameters are presented in Fig. \ref{Fig:LCDM_Redshift}.


\section{Results}\label{Sec:Results}
\subsection{Cosmological posteriors}\label{Sec:Posteriors}
We present our flat \lcdm{} constraints for the pseudo-$C_{\ell}$ analysis of the KiDS-1000 cosmic shear catalogue, as well as combinations with SDSS BAO and RSD (see Sect. \ref{Sec:ExternalData}), and comparisons with Planck 2018 TTTEEE+lowE\footnote{Here we used the \texttt{plik\_lite\_TTTEEE+lowl+lowE} likelihood installed within \texttt{Monte Python 3}.} \citep{Planck2018-cosmology, 2020-PlanckLikelihood2018}, the band-powers constraints for KiDS-1000 obtained by \cite{2020-Asgari-2ptsK1000} and the $3\times 2$pt analysis from \cite{2020-Heymans-KiDS1000-Cosmology}. A summary of the main parameters constrained by cosmic shear data, $\sigma_8$, $\Omega_{\rm m}$, and $S_8$, using MAP+PJ-HPD constraints are presented in Table \ref{table:summary} for KiDS-1000 \pcl{}s and its combination with SDSS BAO and RDS data. A more complete summary of results, including the traditional marginal constraints, can be found in Table \ref{table:FullResults}.  Unless stated otherwise, the quoted credible intervals in this section are the $68\%$ credible intervals (CI) using the MAP+PJ-HPD constraints.

\begin{table}
\caption{Summary of the main parameters constrained by cosmic shear data and combinations with galaxy clustering.} 
\label{table:summary} 
\begin{tabular}{ccc}
\hline\hline
Parameter  & KiDS-1000 P$C_{\ell}$      & KiDS-1000 P$C_{\ell}$ + SDSS  \\ \hline\\[-0.2cm]
$S_8$      & $0.754_{-0.029}^{+0.027}$  &  $0.771_{-0.032}^{+0.006}$    \\[0.3cm] 
$\sigma_8$ & $0.920_{-0.215}^{+0.159}$  &  $0.756_{-0.044}^{+0.029}$    \\[0.3cm] 
$\Omega_{\rm m}$ & $0.202_{-0.053}^{+0.127}$  &  $0.312_{-0.018}^{+0.011}$    \\[0.3cm] 
\hline
\end{tabular}
\tablefoot{Here, we are considering the 68\% C.I. (1$\sigma$) MAP+PJ-PHD. It is important to note that $\sigma_8$ and $\Omega_{\rm m}$ are not properly constrained by cosmic shear alone.}
\end{table}

\begin{figure}
  \centering
  \includegraphics[width=\columnwidth]{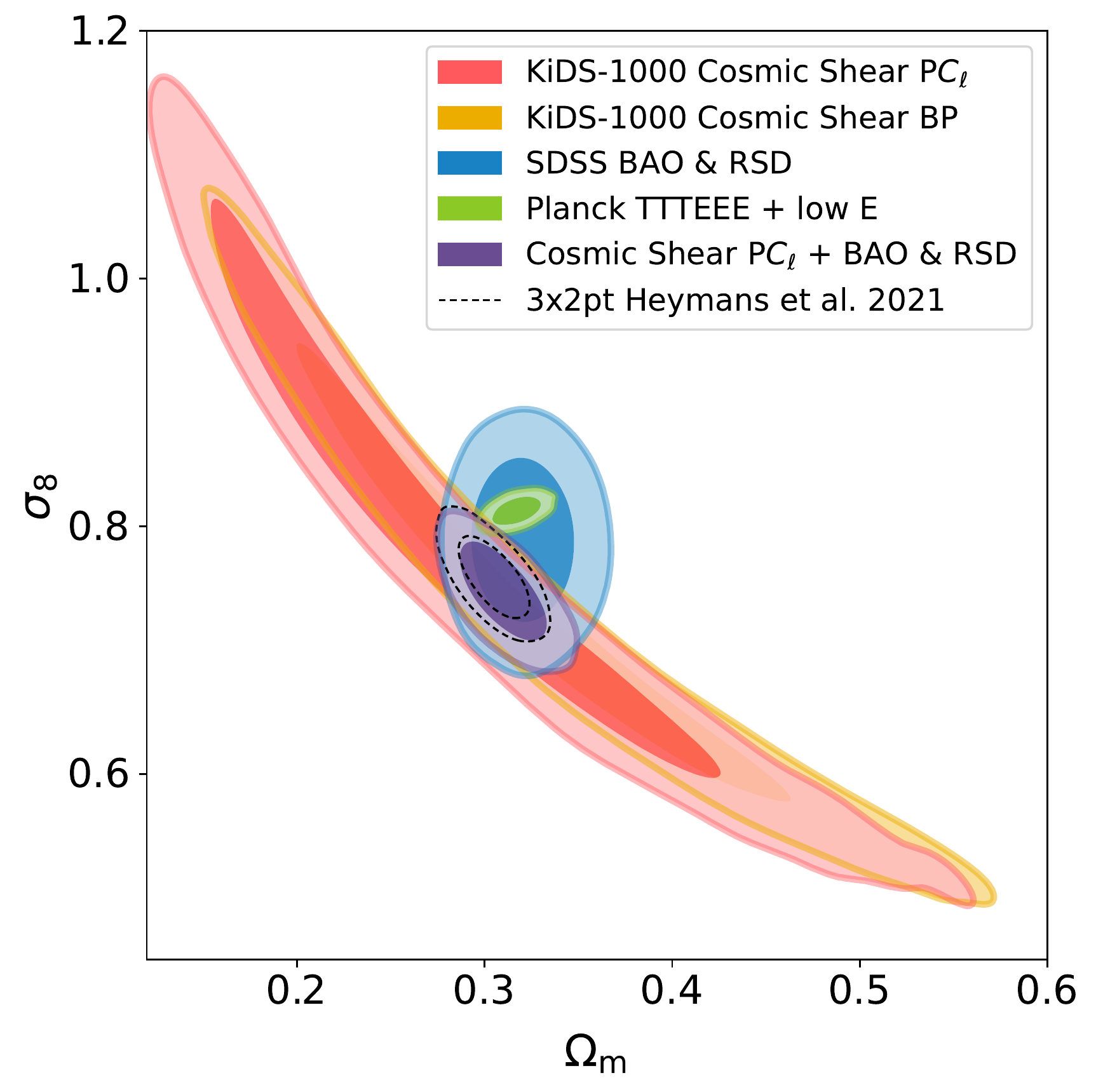}
      \caption{Marginalised 2D posterior distributions for 68\% (darker) and 95\% (lighter) C.I. in the $\Omega_{\rm m} - \sigma_8$ plane. Here we exhibit the \pcl{} constraints (red) in comparison with band-powers results from \cite{2020-Asgari-2ptsK1000} (orange). SDSS BAO and RSD (blue) and its combination with the KiDS-1000 \pcl{} measurements (purple) have a completely different degeneracy than the Planck 2018 constraints from the TTTEEE-lowE combination (green). We also compare it with the $3\times 2$pt analysis from \cite{2020-Heymans-KiDS1000-Cosmology} (black-dashed).}
         \label{Fig:LCDM_Sig8_Om}
\end{figure}

\begin{figure}
  \centering
  \includegraphics[width=\columnwidth]{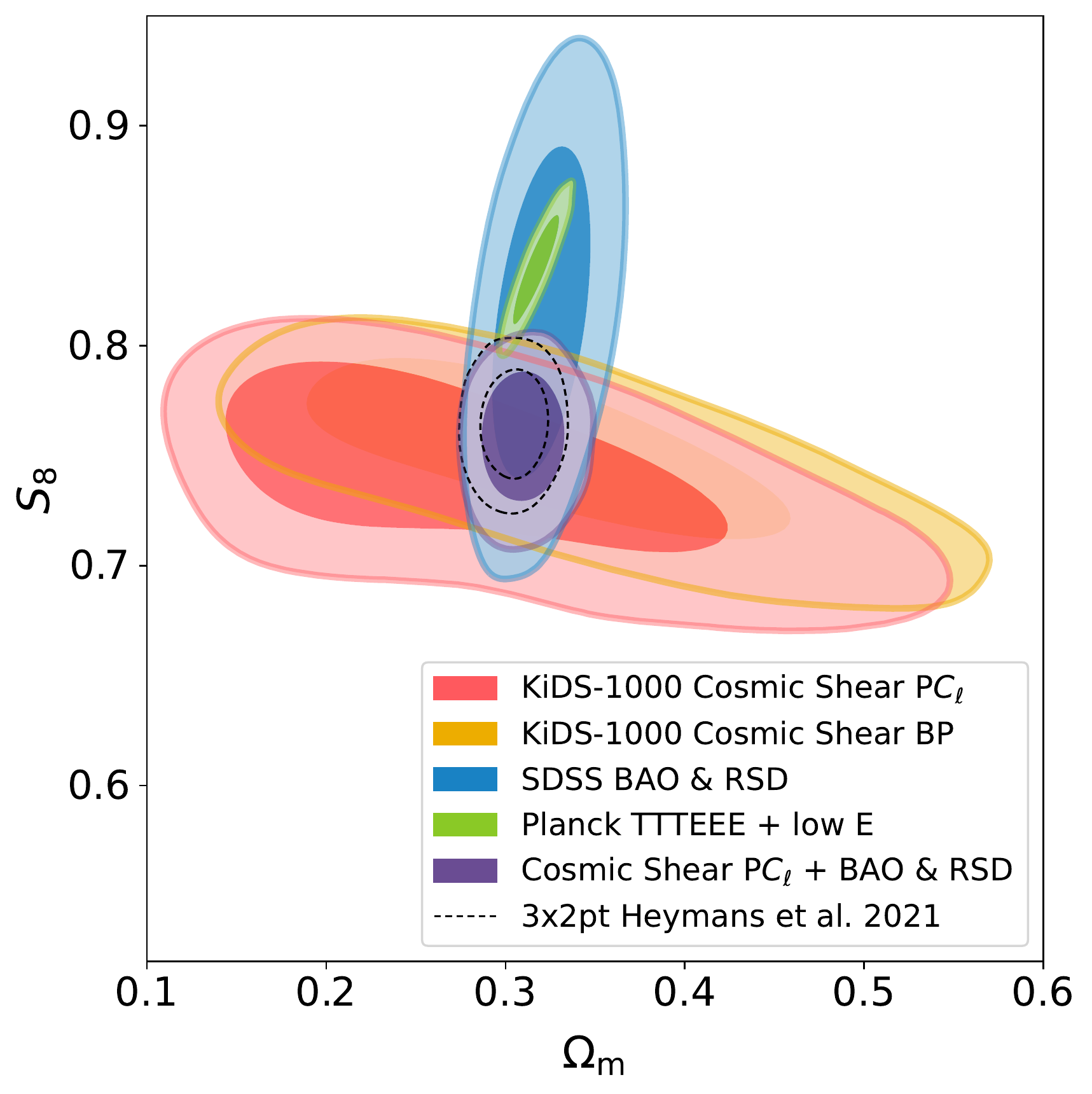}
      \caption{Same as Fig. \ref{Fig:LCDM_Sig8_Om}, but for the marginalised $\Omega_{\rm m}$ and $S_8$ constraints.}
         \label{Fig:LCDM_S8_Om}
\end{figure}

In Fig.  \ref{Fig:LCDM_Sig8_Om}, we display the known cosmic shear degeneracy between the amplitude of matter fluctuations, $\sigma_8$, and the matter density parameters, $\Omega_{\rm m}$. Results between the KiDS-1000 estimators are consistent, with the \pcl{} results (in red) demonstrating a slightly lower value for $\sigma_8$ than the band-powers constraints (in orange) but similar to other 2pt estimators presented in \cite{2020-Asgari-2ptsK1000}. No degeneracy appears for the SDSS BAO and RSD analysis on this parameter plane, meaning that its combination with cosmic shear considerably breaks the known cosmic shear degeneracy on this plane. This is in line with the $3\times 2$pt analysis performed by \cite{2020-Heymans-KiDS1000-Cosmology}, shown as the black dashed contours in Fig. \ref{Fig:LCDM_Sig8_Om}, with the difference that the SDSS BAO and RSD data do not assume flat \lcdm{}, that is it does not use the full galaxy power spectrum, yielding slightly larger constraints. 

The clustering posteriors we obtain demonstrate a slightly higher value of $\sigma_8$ than the BOSS constraints presented in \cite{2020-Heymans-KiDS1000-Cosmology}, which is more apparent when considering the $S_8=\sigma_8\sqrt{\Omega_{\rm m}/0.3}$ parameter instead, as shown in Fig. \ref{Fig:LCDM_SDSS}. Here, we find $S^{\text{SDSS}}_8 = 0.812_{-0.049}^{+0.050}$ for the SDSS BAO and RSD data set which is in agreement with both KiDS-1000 \pcl, $S_8^{\text{PCL}} = 0.754_{-0.029}^{+0.027}$, and the Planck 2018 results. The two main reasons why our clustering constraints differ from those in \cite{2020-Heymans-KiDS1000-Cosmology} are the addition of high-redshift eBOSS LRGs and the Ly-$\alpha$ measurements as well as the impact of not using the full measured galaxy power spectrum for the SDSS measurements, like in \cite{2020-Tilman-BOSS-LCDM}. Yet, when combining the clustering information with cosmic shear, we can see from Fig.~\ref{Fig:LCDM_S8_Om} that the difference is mostly in the $\Omega_{\rm m}$ parameter. This demonstrates the $S_8$ measurement robustness from KiDS-1000 cosmic shear analyses. The $S_8$ constraints with the \pcl{} estimator are slightly larger than the one found with band-powers from \cite{2020-Asgari-2ptsK1000}, reflecting a $3.7\%$ increase in comparison. This is in line with the expected increase due to our noisy covariance matrix estimation method, which applies the estimator to simulations instead of using a theoretical covariance.

A summary of $S_8$ measurements with comparisons to other cosmic shear and CMB probes is displayed in Fig. \ref{Fig:TensionsS8}. Overall, weak lensing experiments are consistent between each other. The purple vertical shaded region in Fig. \ref{Fig:TensionsS8} reflects our main constraints from KiDS-1000 \pcl{} combined with BAO and RSD from BOSS and eBOSS LRGs, and BAO from Ly$\alpha$ forest and its cross-correlations with quasars (SDSS BAO and RSD). Results from the Hyper Supreme Camera (HSC) first year data \citep{2019-Hikage-HSC}, Dark Energy Survey Year 1 (DES-Y1) analysis \citep{2018-Troxel}, and the CFHTLenS constraints from \cite{2017-CHFT-Shahab}\footnote{Here, we quote the \textit{mid} nominal values as this analysis is closer to the analysis performed for KV-450 and KiDS-1000.} are all in agreement with our analysis. Consistency between KiDS-1000 and previous KiDS analyses, such as KV450 \citep{2020-Hidelbrandt-KV450} and KiDS-450 \citep{2017-Hildebrandt-KiDS450}, was explored in detail by \cite{2020-Asgari-2ptsK1000}. A lower value for $S_8$, when compared to the CMB constraints, is consistently found across a range of different and independent cosmic shear experiments carried out by various research groups. 

When comparing the amplitude of intrinsic alignments between the \pcl{} and band-power constraints, we find lower values in the \pcl{} analysis, in line with what has been found for the other 2pt statistics in \cite{2020-Asgari-2ptsK1000}, COSEBIs and correlation functions:

\begin{align}
    \text{[PCL] } &    A_{\text{IA}} = 0.396_{-0.931}^{+0.251}\, ;\\
    \text{[2PCF] }&    A_{\text{IA}} = 0.387_{-0.374}^{+0.321}\, ;\\
    \text{[COSEBIs] }& A_{\text{IA}} = 0.264_{-0.337}^{+0.424}\, ;\\
    \text{[BP] }&      A_{\text{IA}} = 0.973_{-0.383}^{+0.292}\, .
\end{align}
 We speculate that this could be the reason why our method finds a slightly lower peak for the matter density parameter, $\Omega_{\rm m}$, as a similar effect appears in \cite{2020-Asgari-2ptsK1000}. Despite this, $\Omega_{\rm m}$ is not strongly constrained in our analysis, meaning that the uncertainty is large enough that the constraints remain consistent for both $\Omega_{\rm m}$ and $A_{\text{IA}}$, as investigated by \cite{2020-Asgari-2ptsK1000}. Since both band-powers and \pcl{}s are in harmonic space, one could expect them to match better than \pcl{}s and correlation functions, but the difference between both measurements is $\sim 1.2\sigma$ away (see Eq. \ref{Eq:TensionMetricsTau}). We note that in contrast to the other 2pt functions presented in \cite{2020-Asgari-2ptsK1000}, band-powers filter out most of the scales below $\ell = 100$. Meanwhile, correlation functions and COSEBIs have filters in harmonic space that capture power below $\ell = 100$. Comparing this with the \pcl{}s $\ell_{\text{min}} = 76$, Fig.~\ref{Fig:Windows} shows that larger scales also leak through the mixing matrix filters into our analysis. This slight difference in $\ell$ range potentially explains the differences between band-powers and \pcl, and their similarity with the correlation function and COSEBIs results since intrinsic alignments have a significant impact in larger scales as seen in Fig. 4 from \cite{2020-Asgari-2ptsK1000}.

The full posteriors for cosmological parameters and the lensing nuisance parameters, $A_{\rm bary}$ and $A_{\text{IA}}$, are shown in Fig.~\ref{Fig:LCDM_Cosmo}, and the redshift nuisance parameters are shown in the Appendix in Fig.~\ref{Fig:LCDM_Redshift} with a comparison to the band-power constraints from \cite{2020-Asgari-2ptsK1000}. Table~\ref{table:FullResults} presents a summary of our cosmological inference results, including the traditional marginalised constraints. For our best-fit parameters, the reduced $\chi^2$ for the cosmic shear-only analysis is 1.18, reflecting an acceptable \textit{p}-value of 0.06.


\subsection{Cosmic shear tension with early Universe probes}
\begin{figure}
  \centering
  \includegraphics[width=\columnwidth]{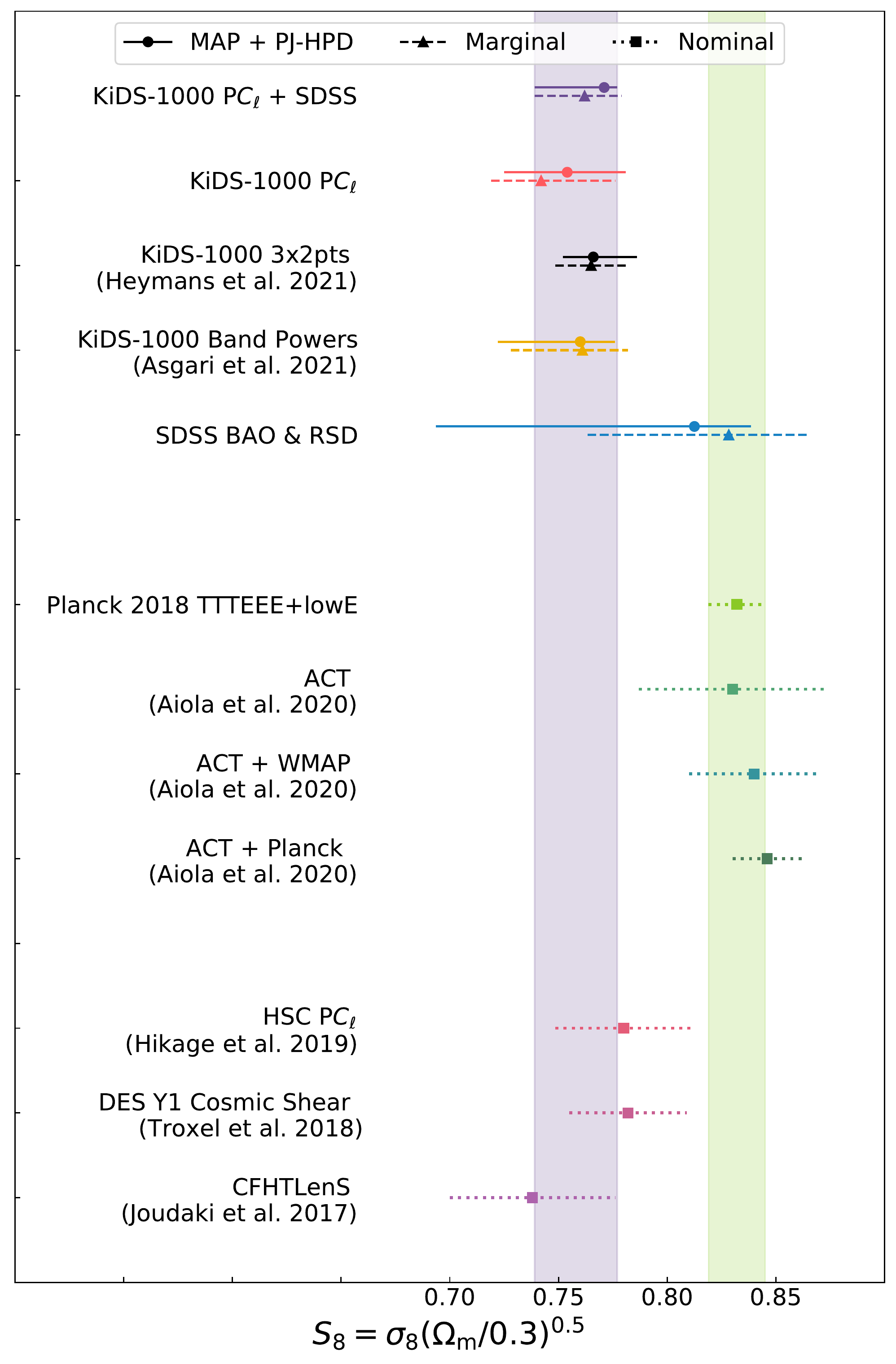}
  
      \caption{Comparison between different cosmological probe constraints on the growth of structure parameter, $S_8 = \sigma_8(\Omega_{\rm m}/0.3)^{1/2}$. Here, we present our results with both the fiducial MAP+PJ-HPD constraints (solid) and the traditional marginal posterior mode (dashed). The purple shaded region reflects the MAP+PJ-HPD constraints for KiDS-1000 \pcl{} plus SDSS BAO and RSD. For external probes quoted from the literature, we quote the marginal posterior mode with tail credible intervals (dotted). Our analysis is not only consistent with other KiDS-1000 analysis but also consistent with other independent cosmic shear probes. Tensions between cosmic shear experiments and Planck 2018 (green shaded region) or ACT+Planck are between 2.8 to 3.2$\sigma$. }
         \label{Fig:TensionsS8}
\end{figure}

\begin{figure*}
   \centering
   \includegraphics[width=2\columnwidth]{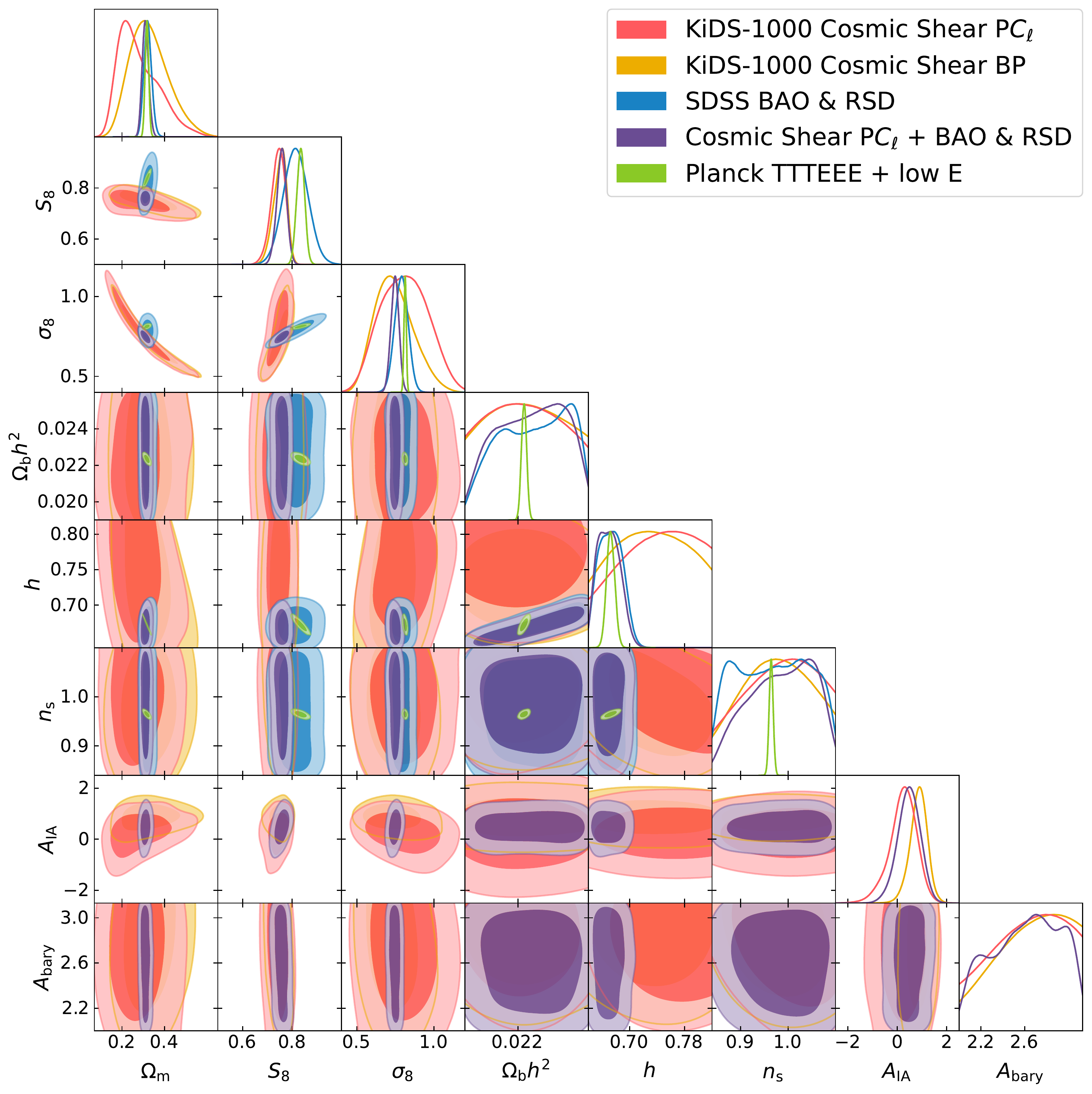}
      \caption{Extended marginalised 1D and 2D posteriors for relevant \lcdm{} and astrophysical parameters: the matter density parameter ($\Omega_{\rm m}$), the amplitude of matter density fluctuations ($S_8 = \sigma_8\sqrt{\Omega_{\rm m}/0.3}$),  $\sigma_8$ as a derived parameter, the physical baryonic matter density ($\Omega_{\rm b} h^2$), the dimensionless Hubble constant ($h$), and the spectral index ($n_{\rm s}$).  Included are also two nuisance parameters: the amplitude of intrinsic alignments ($A_{\rm IA}$) and the baryonic feedback amplitude ($A_{\rm bary}$) as these correlate with some cosmological parameters. When combining the KiDS-1000 \pcl{} constraints (red) with the SDSS BAO and RSD constraints (blue), we obtain constraints (purple) that break a few degeneracies that cosmic shear alone cannot. For comparison, we added the Planck 2018 TTTEEE-lowE \citep{2020-PlanckLikelihood2018} constrains (green) and the KiDS-1000 band-powers constraints (orange) from \cite{2020-Asgari-2ptsK1000}. }
         \label{Fig:LCDM_Cosmo}
\end{figure*}
We now turn our discussion towards the known inconsistencies between cosmic shear surveys and the cosmic microwave background in the amplitude of matter density fluctuations and, consequently, the growth of structure as measured by $S_8$. By assuming Gaussianity and adopting the marginal constraints to quantify tension, we initially employ the conventional marginal tension estimate

\begin{align}
    \tau = \frac{|\langle S_8^{\rm A} \rangle - \langle S_8^{\rm B} \rangle|}{\sqrt{\sigma^2(S_8^{\rm A})+\sigma^2(S_8^{\rm B})}}\, ,
    \label{Eq:TensionMetricsTau}
\end{align}
where $\langle S_8^{\rm A} \rangle$ and $\sigma^2(S_8^{\rm A})$ are the mean and variance for a given probe A. Under the Gaussianity assumption, $\tau$ measures how consistent the difference between $\langle S_8^{\rm A} \rangle$ and $\langle S_8^{\rm B} \rangle$ is with a case where there is no difference between measurements.

Comparisons between KiDS-1000 \pcl{} measurements and Planck 2018 TTTEEE+lowE nominal results \citep{Planck2018-cosmology, 2020-PlanckLikelihood2018} lead to a $2.8\sigma$ tension between late- and early-Universe constraints. When combining the \pcl{} measurements with SDSS BAO and RSD data, the tension with Planck 2018 increases to 2.9$\sigma$, in line with the tension found in the KiDS-1000 $3\times 2$pt analysis from \cite{2020-Heymans-KiDS1000-Cosmology}. 

To further assess the tension between late-time and early-time probes, we verify the tension between our KiDS-1000 \pcl{} constraints and the most recent results from the Atacama Cosmology Telescope (ACT), a ground-based CMB experiment \citep{2020-ACT-1}. We compare our results with ACT alone, ACT combined with WMAP, and ACT combined with Planck 2018.  These nominal constraints can also be seen in Fig.~\ref{Fig:TensionsS8}. 

As expected, tensions with ACT constraints alone are much smaller given that the data set is not as precise as space-based CMB observations; we find $\tau = 1.4\sigma$ when combining cosmic shear and clustering data. However, ACT does not measure the TT power spectrum below $\ell =600$, and the TE and EE power spectra below multipole 350, losing the first two acoustic peaks in the temperature auto spectrum and the entirety of the first peak in the TE/EE spectra. As a consequence, ACT's nominal constraints for $S_8$ include information from the final WMAP public data release \citep{2013-Bennet-WMAP}. The tension between our constraints from cosmic shear and clustering increases to $2.2\sigma$ for the nominal ACT plus WMAP results. 

The largest tension between cosmic microwave background and large-scale structure measurements appears when combining {ACT} and \textit{Planck}. For \pcl{} alone, we find $3.2\sigma$ tension and combining cosmic shear with clustering, we find $3.3\sigma$ tension. However, using the suspiciousness statistic \citep{2020-Lemos-Susp}, \cite{2020-Handley-Lemos-CMB-Tension} pointed out that Planck 2018 and ACT have a $2.6\sigma$ tension between each other.

It is clear from Fig. \ref{Fig:LCDM_Cosmo} and Table \ref{table:summary}, that the marginal distribution of $S_8$ is not Gaussian. To further assess the tension between our KiDS-1000 \pcl{} measurements and the Planck Legacy results, we also adopt the Hellinger distance measure, $d_{\text{H}}$, as a tension metric (\citealt{1977-Beran-HellingerTension}, and more details in Appendix F from \citealt{2020-Heymans-KiDS1000-Cosmology}). The Hellinger distance is a stable metric for the comparison of arbitrary one-dimensional distributions. For two distinct distributions, $p(\theta)$ and $q(\theta$), $d_{\text{H}}$ is defined as

\begin{align}
    d_{\text{H}}^2[p, q] = \frac{1}{2}\int {\rm d}\theta \left[ \sqrt{p(\theta)} - \sqrt{q(\theta}) \right]^2\, .
\end{align}
With the metric above, we measured $ d_{\text{H}} = 0.95$ between our KiDS-1000 \pcl{} and Planck Legacy's $S_8$ constraints. Using the methodology described by \cite{2020-Heymans-KiDS1000-Cosmology}, this can be expressed as a $\sim 3.1\sigma$ tension. For our combination of cosmic shear from KiDS and clustering from SDSS, we obtain  $ d_{\text{H}} = 0.94$ -- corresponding to a $\sim 3\sigma$ tension with \textit{Planck}. Since this tension metric uses the marginalised distribution for a given parameters, one needs the MCMC chains to calculate it. The ACT DR4 chains were not publicly available at the time. Therefore, we were not able to verify $d_{\text{H}}$ between KiDS-1000, ACT and combinations with ACT.


\section{Conclusions}\label{Sec:Conclusions}
We performed a pseudo angular power spectra (\pcl) analysis of the current state-of-the-art  galaxy shape measurements from the Kilo Degree Survey's fourth data release (KiDS-1000, \citealt{2019-KiDS-DR4}), obtaining cosmological constraints for the current concordance model, the flat \lcdm{} cosmology. The KiDS-1000 shear catalogue \citep{2020-KiDS-1000-ShearCat} is now publicly available\footnote{\url{http://kids.strw.leidenuniv.nl/DR4/KiDS-1000_shearcatalogue.php} \newline \url{https://github.com/KiDS-WL/Cat_to_Obs_K1000_P1}}, containing nine-band photometry with shape and photometric redshift estimates for more than $21$ million lensed objects distributed over an area of around 1000 deg$^2$.

With this work, we added an additional 2pt function to the family of measurements used in \cite{2020-Asgari-2ptsK1000}. With its highly accurate calibrated shear and photometric redshift measurements, the KiDS-1000 shear catalogue is an excellent proxy for future surveys, including their early data-releases. The performance of our \pcl{} estimator on KiDS-1000 data marks an important road-test for its future application on data from \textit{Euclid} cosmic shear catalogue. Even for a comparatively small survey area and a disjointed footprint, the \pcl{} estimator performed as well as other 2pt functions, with no flat-sky approximation --- an important and necessary characteristic  that estimators of the next generation of galaxy surveys need. The forward-modelling approach allowed us to maintain accuracy and avoid numerical instabilities from inverting and deconvolving a mixing matrix from our estimates with no significant computational cost when calculating our theory-vector for cosmological inference. 

{Through this re-analysis in harmonic space of KiDS-1000 data as a road-test of our prototype estimator for the \textit{Euclid} Science Ground Segment we find that, with KiDS data, we are operating at the limit of this methodology. First, noise realisations of the galaxy shape catalogue will not be feasible for the size of the expected \textit{Euclid} shape catalogue ($\sim 2\times 10^9$ galaxies); hence we will implement a noise power spectra subtraction based on the theoretical expectation of the noise. Second, the treatment used for the shear weights in the analysis is sufficient for KiDS, but it may not be the case for \textit{Euclid}. For higher resolution shear maps such as those required by the \textit{Euclid} Science Ground Segment ($N_{\rm side}=4096$), the shear weights will need to be incorporated into the mask, requiring our formalism to change. Finally, we will also investigate the possibility of deconvolving the effects of the mixing matrix while maintaining the level of accuracy and precision required by \textit{Euclid} outlined in \cite{2011Euclid}, which does not allow for mask apodisation. We have learned valuable lessons for \textit{Euclid} which we will explore in a subsequent paper (Euclid Consortium et al., \textit{in prep}).}

Following, our systematic contamination null-test analysis also used the \pcl{} estimator at its core, demonstrating the versatility of this method. Using 2pt statistics in harmonic space to assess systematic contamination yields insights regarding the validity of cosmological information on the scales used to infer cosmology. This approach allows us to be confident when using a broad range of multipoles from measured angular power spectra as any spurious correlations arising solely from systematic contamination will be picked-up by it. We find no significant correlations between systematics and data for any relevant observational catalogue quantities nor for stellar density from Gaia DR2. These null detection analyses, together with the null detection of B-modes in our data, made us confident that the measured E-mode \pcl{}s contained only cosmological and astrophysical signals and could be used for Bayesian inference of the relevant flat \lcdm{} parameters.

Cosmic shear alone constrains the growth of structure parameter, $S_8$, quite well but it has a known degeneracy in the amplitude of matter density fluctuations and density of matter plane of the parameter space. To break this degeneracy, we combined our \pcl{} measurements with baryon acoustic oscillations and redshift space distortion measurements from luminous red galaxies from the Sloan Digital Sky Survey's BOSS DR12 and eBOSS DR16 data sets, as well as baryon acoustic oscillations from the Lyman-$\alpha$ forest and its cross-correlations with quasar positions from eBOSS DR16 \citep{2020-eBOSSDR16}. The addition of clustering data allowed us to break the aforementioned degeneracy. 

Our cosmological inference takes a few cosmic shear and redshift distribution-related nuisance parameters into account. From these, we can observe an interesting trend for the amplitude of intrinsic alignments, $A_{\text{IA}}$, when compared to previous results from \cite{2020-Asgari-2ptsK1000}. In contrast to band-power measurements, \pcl{}, as well as correlation functions and COSEBIs, found a lower value of $A_{\text{IA}}$. We believe that these results are related to the larger scales accessible by the different analyses. Band-powers have a higher maximum scale cut at $\ell_{\text{min}}=100$, while we used $\ell_{\text{min}}=76$ for \pcl{}, for example. We emphasise that, given the uncertainty in these measurements, there is no evidence for inconsistencies from $A_{\text{IA}}$ between the different 2pt functions as measurements agree up to $\sim 1.2\sigma$ (using the same metrics as in Eq. \ref{Eq:TensionMetricsTau}).

Quoting the {maximum a posteriori} (MAP) value with the projected joint highest posterior density (PJ-HPD) credible intervals \citep{2020-KiDS1000-Methods}, our \pcl{} analysis alone yielded $S_8 = 0.754_{-0.029}^{+0.027}$, or $9.8\pm 3.3\%$ lower than the $S_8$ nominal constraints from \cite{Planck2018-cosmology}. When adding clustering data from SDSS, we increased the precision on the growth of structure parameter, obtaining a measurement of $S_8 = 0.771^{+0.006}_{-0.032}$, reflecting a $7.8\pm2.3\%$ lower value when comparing the marginal constraints with \textit{Planck}'s measurement of $S_8$. Both constraints are in very good agreement with other KiDS-1000 measurements from \cite{2020-Asgari-2ptsK1000,2020-Heymans-KiDS1000-Cosmology} and other cosmic shear surveys such as the Dark Energy Survey \citep{2018-Troxel}, the Hyper Supreme Camera survey \citep{2019-Hikage-HSC} and CFHTLenS \citep{2017-CHFT-Shahab}. 

Using two different metrics to evaluate tension in a marginalised space, we assessed the tension in the growth of structure parameter between cosmic shear and probes of the cosmic microwave background. Tension with the Planck Legacy TTTEEElowE analysis is around 2.8$\sigma$ for cosmic shear alone, in line with what was found by \cite{2020-Asgari-2ptsK1000}. When comparing the Planck Legacy results to our combination of cosmic shear with clustering data from SDSS, the tension increased to $\sim 2.9\sigma$, also in line with findings from \cite{2020-Heymans-KiDS1000-Cosmology}. 

We further verify that the tensions we observed did not arise exclusively from comparisons with the Planck Mission results and there is indeed a tension between the large-scale structure of the late Universe and the cosmic microwave background radiation from the early Universe. We analysed this by comparing our results to the nominal constraints from a ground-based CMB experiment, the Atacama Cosmology Telescope Data Release 4 (ACT, \citealt{2020-ACT-1}). The nominal results from ACT also included data from nine years of observations from WMAP \citep{2013-Bennet-WMAP}; the tension between our constraints from cosmic shear combined with clustering and ACT+WMAP were not as high as with Planck Legacy: $2.2\sigma$. However, ACT was also combined with \textit{Planck} instead of WMAP. In this case, tensions between large-scale structure probes and the cosmic microwave background were as high as $3.3\sigma$. The impact of \textit{Planck} on this tension value, however apparent, cannot be properly assessed as evidence points to internal tensions between ACT, WMAP, and \textit{Planck} \citep{2020-Handley-Lemos-CMB-Tension}. The tensions between the late and early Universe seem to be independent of {Planck} Legacy data, but are exacerbated by it.

So far, the tension has remained at a level that is inconclusive as to whether it is arising from unknown systematics in the data, noise in the realisation of the portions of the Universe that we are observing, or hints of new physics. The Kilo Degree Survey still has a final future data release (DR5) and our hopes are that, with it, we will obtain a better understanding of what is causing this tension with the growth of structure in the Universe. Nonetheless, the next decade will see the light of new and more powerful cosmic shear surveys with the launch of the Euclid Space Telescope and the start of the survey carried out by the Vera C. Rubin Observatory. Both surveys, independently and combined, will produce an almost complete sky map of the large-scale structure of the Universe, using cosmic shear as one of their main probes of understanding cosmology, and with the potential of finally solving this cosmic puzzle.


\begin{acknowledgements}
The authors would like to thank Dr Lorne Whiteway for useful comments and discussions. We would also like to thank the reviewer for their helpful comments that helped enrich this work. The mixing matrix calculation used a publicly available Wigner 3-j symbol code developed by Dr Whiteway and it can be found in \url{https://github.com/LorneWhiteway/UCLWig3j}. The K-Means code used for the jackknife covariances for systematics was built on top of a code developed by Dr Erin Sheldon, publicly available in \url{https://github.com/esheldon/kmeans_radec}. The analysis performed in this work also made use of the following packages and software: \texttt{NumPy} \citep{harris2020numpy}, \texttt{SciPy} \citep{2020SciPy-NMeth}, \texttt{Matplotlib} \citep{Hunter:2007-matplotlib}, \texttt{GetDist} \citep{2019-Lewis-GetDist}, \texttt{Pandas} \citep{mckinney-proc-scipy-2010}, \texttt{AstroPy} \citep{astropy:2018} and \texttt{Jupyter Lab} \citep{jupyter}. The \flasmo{} mocks were generated and provided by Dr Chieh-An Lin. We thank Dr Edd Edmonson for technical support at the UCL HPC facilities. This work used computing equipment funded by the Research Capital Investment Fund (RCIF) provided by UKRI, and is partially funded by the UCL Cosmoparticle Initiative. MA and TT acknowledge support from the European Research Council under grant number 647112. AHW and AD acknowledge support from the European Research Council Consolidator Grant (No. 770935). MB is supported by the Polish National Science Center through grants no. 2020/38/E/ST9/00395, 2018/30/E/ST9/00698 and 2018/31/G/ST9/03388, and by the Polish Ministry of Science and Higher Education through grant DIR/WK/2018/12. BG acknowledges support from the European Research Council under grant number 647112 and from the Royal Society through an Enhancement Award (RGF/EA/181006). CH acknowledges support from the European Research Council under grant number 647112, and support from the Max Planck Society and the Alexander von Humboldt Foundation in the framework of the Max Planck-Humboldt Research Award endowed by the Federal Ministry of Education and Research. H. Hildebrandt is supported by a Heisenberg grant of the Deutsche Forschungsgemeinschaft (Hi 1495/5-1) as well as an ERC Consolidator Grant (No. 770935). HYS acknowledges the support from NSFC of China under grant 11973070, the Shanghai Committee of Science and Technology grant No.19ZR1466600 and Key Research Program of Frontier Sciences, CAS, Grant No. ZDBS-LY-7013. \newline
The  KiDS-1000  data products  in  this  paper  are  based on observations made with ESO Telescopes at the La Silla Paranal Observatory under programme IDs 177.A-3016, 177.A-3017 and 177.A-3018, and on data products produced by Target/OmegaCEN, INAF-OACN, INAF-OAPD and the KiDS production team, on behalf of the KiDS consortium. \newline
\AckECon
\newline
\textit{Author Contributions:} All authors contributed to the development and writing of this paper. The authorship list is given in several groups: the lead authors (AL, LW, ASM, BJ, AC), followed by an alphabetical group that includes those who are key contributors to both the scientific analysis and the KiDS data products, and a further alphabetical group covering those who have either made a significant contribution to the KiDS data products or to the scientific analysis. Additional alphabetical groups correspond to different authorship levels of the Euclid Consortium.

\end{acknowledgements}

\bibliographystyle{aa}
\bibliography{bibliogr}

\section*{Affiliations}
{\footnotesize
$^1$ Department of Physics and Astronomy, University College London, Gower Street, London WC1E 6BT, UK\\
$^2$ Astrophysics Group, Blackett Laboratory, Imperial College London, London SW7 2AZ, UK\\
$^3$ Institute for Astronomy, University of Edinburgh, Royal Observatory, Blackford Hill, Edinburgh EH9 3HJ, UK\\
$^4$ Jodrell Bank Centre for Astrophysics, Department of Physics and Astronomy, University of Manchester, Oxford Road, Manchester M13 9PL, UK\\
$^5$ Mullard Space Science Laboratory, University College London, Holmbury St Mary, Dorking, Surrey RH5 6NT, UK\\
$^6$ E. A. Milne Centre, University of Hull, Cottingham Road, Hull, HU6 7RX, UK\\
$^7$ Ruhr-Universit\"at Bochum, Astronomisches Institut, German Centre for Cosmological Lensing (GCCL), Universit\"atsstr. 150, 44801 Bochum, Germany\\
$^8$ Center for Theoretical Physics, Polish Academy of Sciences, al. Lotnik\'{o}w 32/46, 02-668 Warsaw, Poland\\
$^9$ Astronomisches Institut, Ruhr-Universit\"at Bochum, Universit\"atsstr. 150, 44801 Bochum, Germany\\
$^10$ Shanghai Astronomical Observatory (SHAO), Nandan Road 80, Shanghai 200030, China\\
$^11$ University of Chinese Academy of Sciences, Beijing 100049, China\\
$^12$ Institute of Cosmology and Gravitation, University of Portsmouth, Portsmouth PO1 3FX, UK\\
$^13$ INAF-Osservatorio di Astrofisica e Scienza dello Spazio di Bologna, Via Piero Gobetti 93/3, I-40129 Bologna, Italy\\
$^14$ Max Planck Institute for Extraterrestrial Physics, Giessenbachstr. 1, D-85748 Garching, Germany\\
$^15$ INAF-Osservatorio Astrofisico di Torino, Via Osservatorio 20, I-10025 Pino Torinese (TO), Italy\\
$^16$ Department of Mathematics and Physics, Roma Tre University, Via della Vasca Navale 84, I-00146 Rome, Italy\\
$^17$ INFN-Sezione di Roma Tre, Via della Vasca Navale 84, I-00146, Roma, Italy\\
$^18$ INAF-Osservatorio Astronomico di Capodimonte, Via Moiariello 16, I-80131 Napoli, Italy\\
$^19$ INAF-IASF Milano, Via Alfonso Corti 12, I-20133 Milano, Italy\\
$^20$ Institut de F\'{i}sica d'Altes Energies (IFAE), The Barcelona Institute of Science and Technology, Campus UAB, 08193 Bellaterra (Barcelona), Spain\\
$^21$ Port d'Informaci\'{o} Cient\'{i}fica, Campus UAB, C. Albareda s/n, 08193 Bellaterra (Barcelona), Spain\\
$^22$ INAF-Osservatorio Astronomico di Roma, Via Frascati 33, I-00078 Monteporzio Catone, Italy\\
$^23$ Department of Physics "E. Pancini", University Federico II, Via Cinthia 6, I-80126, Napoli, Italy\\
$^24$ INFN section of Naples, Via Cinthia 6, I-80126, Napoli, Italy\\
$^25$ Dipartimento di Fisica e Astronomia "Augusto Righi" - Alma Mater Studiorum Universit\'a di Bologna, Viale Berti Pichat 6/2, I-40127 Bologna, Italy\\
$^26$ INAF-Osservatorio Astrofisico di Arcetri, Largo E. Fermi 5, I-50125, Firenze, Italy\\
$^27$ Centre National d'Etudes Spatiales, Toulouse, France\\
$^28$ Institut national de physique nucl\'eaire et de physique des particules, 3 rue Michel-Ange, 75794 Paris C\'edex 16, France\\
$^29$ European Space Agency/ESRIN, Largo Galileo Galilei 1, 00044 Frascati, Roma, Italy\\
$^30$ ESAC/ESA, Camino Bajo del Castillo, s/n., Urb. Villafranca del Castillo, 28692 Villanueva de la Ca\~nada, Madrid, Spain\\
$^31$ Univ Lyon, Univ Claude Bernard Lyon 1, CNRS/IN2P3, IP2I Lyon, UMR 5822, F-69622, Villeurbanne, France\\
$^32$ Departamento de F\'isica, Faculdade de Ci\^encias, Universidade de Lisboa, Edif\'icio C8, Campo Grande, PT1749-016 Lisboa, Portugal\\
$^33$ Instituto de Astrof\'isica e Ci\^encias do Espa\c{c}o, Faculdade de Ci\^encias, Universidade de Lisboa, Campo Grande, PT-1749-016 Lisboa, Portugal\\
$^34$ Universit\'e Paris-Saclay, CNRS, Institut d'astrophysique spatiale, 91405, Orsay, France\\
$^35$ Department of Astronomy, University of Geneva, ch. d\'Ecogia 16, CH-1290 Versoix, Switzerland\\
$^36$ Department of Physics, Oxford University, Keble Road, Oxford OX1 3RH, UK\\
$^37$ INFN-Padova, Via Marzolo 8, I-35131 Padova, Italy\\
$^38$ AIM, CEA, CNRS, Universit\'{e} Paris-Saclay, Universit\'{e} de Paris, F-91191 Gif-sur-Yvette, France\\
$^39$ Institut d'Estudis Espacials de Catalunya (IEEC), Carrer Gran Capit\'a 2-4, 08034 Barcelona, Spain\\
$^40$ Institute of Space Sciences (ICE, CSIC), Campus UAB, Carrer de Can Magrans, s/n, 08193 Barcelona, Spain\\
$^41$ INAF-Osservatorio Astronomico di Trieste, Via G. B. Tiepolo 11, I-34131 Trieste, Italy\\
$^42$ Istituto Nazionale di Astrofisica (INAF) - Osservatorio di Astrofisica e Scienza dello Spazio (OAS), Via Gobetti 93/3, I-40127 Bologna, Italy\\
$^43$ Istituto Nazionale di Fisica Nucleare, Sezione di Bologna, Via Irnerio 46, I-40126 Bologna, Italy\\
$^44$ INAF-Osservatorio Astronomico di Padova, Via dell'Osservatorio 5, I-35122 Padova, Italy\\
$^45$ Universit\"ats-Sternwarte M\"unchen, Fakult\"at f\"ur Physik, Ludwig-Maximilians-Universit\"at M\"unchen, Scheinerstrasse 1, 81679 M\"unchen, Germany\\
$^46$ Institute of Theoretical Astrophysics, University of Oslo, P.O. Box 1029 Blindern, N-0315 Oslo, Norway\\
$^47$ Jet Propulsion Laboratory, California Institute of Technology, 4800 Oak Grove Drive, Pasadena, CA, 91109, USA\\
$^48$ von Hoerner \& Sulger GmbH, Schlo{\ss}Platz 8, D-68723 Schwetzingen, Germany\\
$^49$ Max-Planck-Institut f\"ur Astronomie, K\"onigstuhl 17, D-69117 Heidelberg, Germany\\
$^50$ Aix-Marseille Univ, CNRS/IN2P3, CPPM, Marseille, France\\
$^51$ Leiden Observatory, Leiden University, Niels Bohrweg 2, 2333 CA Leiden, The Netherlands\\
$^52$ Universit\'e de Gen\`eve, D\'epartement de Physique Th\'eorique and Centre for Astroparticle Physics, 24 quai Ernest-Ansermet, CH-1211 Gen\`eve 4, Switzerland\\
$^53$ Department of Physics and Helsinki Institute of Physics, Gustaf H\"allstr\"omin katu 2, 00014 University of Helsinki, Finland\\
$^54$ NOVA optical infrared instrumentation group at ASTRON, Oude Hoogeveensedijk 4, 7991PD, Dwingeloo, The Netherlands\\
$^55$ Argelander-Institut f\"ur Astronomie, Universit\"at Bonn, Auf dem H\"ugel 71, 53121 Bonn, Germany\\
$^56$ INFN-Sezione di Bologna, Viale Berti Pichat 6/2, I-40127 Bologna, Italy\\
$^57$ Dipartimento di Fisica e Astronomia "Augusto Righi" - Alma Mater Studiorum Universit\`{a} di Bologna, via Piero Gobetti 93/2, I-40129 Bologna, Italy\\
$^58$ Centre for Extragalactic Astronomy, Department of Physics, Durham University, South Road, Durham, DH1 3LE, UK\\
$^59$ Institute for Computational Cosmology, Department of Physics, Durham University, South Road, Durham, DH1 3LE, UK\\
$^60$ Observatoire de Sauverny, Ecole Polytechnique F\'ed\'erale de Lau- sanne, CH-1290 Versoix, Switzerland\\
$^61$ European Space Agency/ESTEC, Keplerlaan 1, 2201 AZ Noordwijk, The Netherlands\\
$^62$ Department of Physics and Astronomy, University of Aarhus, Ny Munkegade 120, DK-8000 Aarhus C, Denmark\\
$^63$ Institute of Space Science, Bucharest, Ro-077125, Romania\\
$^64$ Max-Planck-Institut f\"ur Astrophysik, Karl-Schwarzschild Str. 1, 85741 Garching, Germany\\
$^65$ Dipartimento di Fisica e Astronomia "G.Galilei", Universit\'a di Padova, Via Marzolo 8, I-35131 Padova, Italy\\
$^66$ Centro de Investigaciones Energ\'eticas, Medioambientales y Tecnol\'ogicas (CIEMAT), Avenida Complutense 40, 28040 Madrid, Spain\\
$^67$ Instituto de Astrof\'isica e Ci\^encias do Espa\c{c}o, Faculdade de Ci\^encias, Universidade de Lisboa, Tapada da Ajuda, PT-1349-018 Lisboa, Portugal\\
$^68$ Universidad Polit\'ecnica de Cartagena, Departamento de Electr\'onica y Tecnolog\'ia de Computadoras, 30202 Cartagena, Spain\\
$^69$ Kapteyn Astronomical Institute, University of Groningen, PO Box 800, 9700 AV Groningen, The Netherlands\\
$^70$ Infrared Processing and Analysis Center, California Institute of Technology, Pasadena, CA 91125, USA\\
$^71$ INAF-Osservatorio Astronomico di Brera, Via Brera 28, I-20122 Milano, Italy\\
$^72$ Dipartimento di Fisica, Universit\'a degli Studi di Torino, Via P. Giuria 1, I-10125 Torino, Italy\\
$^73$ INFN-Sezione di Torino, Via P. Giuria 1, I-10125 Torino, Italy\\
$^74$ INAF-IASF Bologna, Via Piero Gobetti 101, I-40129 Bologna, Italy\\
$^75$ Space Science Data Center, Italian Space Agency, via del Politecnico snc, 00133 Roma, Italy}

\appendix
\section{Extra figures and tables}\label{Apdx:Extra}

\begin{table*}
\caption{Details on each tomographic bin.}             
\label{table:1}      
\centering          
\begin{tabular}{c c c r c c}  
\hline\hline       
                    
Bin & $z_{\rm B}$ range & mean $z$ & Number of objects &  $n_{\text{eff}} [\text{arcmin}^{-2}]$ & $m$-correction\\
\hline                    
   1 & $0.1 < z_{\rm B} \leq 0.3$& 0.26 &  1 792 136    & 0.62 & $-0.009\pm 0.019$\\  
   2 & $0.3 < z_{\rm B} \leq 0.5$& 0.40 &  3 681 319    & 1.18 & $-0.011\pm 0.020$\\
   3 & $0.5 < z_{\rm B} \leq 0.7$& 0.56 &  6 148 102    & 1.85 & $-0.015\pm 0.017$\\
   4 & $0.7 < z_{\rm B} \leq 0.9$& 0.79 &  4 544 395    & 1.26 &  $0.002\pm 0.012$\\
   5 & $0.9 < z_{\rm B} \leq 1.2$& 0.98 &  5 096 059    & 1.31 &  $0.007\pm 0.010$\\
\hline                  
total & $0.1 < z_{\rm B} \leq 1.2$& N/A & 21 262 011 & 6.95 & N/A \\
\hline
\end{tabular}
\tablefoot{The effective number of galaxies, $n_{\text{eff}}$, is defined in the same way as in \citealt{2012-Heymans}. Details on the \textit{m}-corrections estimation can be found in \cite{2020-KiDS-1000-ShearCat}.}
\end{table*}

\begin{table*}
\centering
\caption{Summary of marginalised parameter constraints with 68\% confidence intervals. Priors and parameter definitions are shown in Table \ref{table:priors}.} 
\label{table:FullResults}
    \begin{tabular}{lcccc}
        \hline\hline
        Parameter         & KiDS-1000 PCL              & KiDS-1000 PCL         & KiDS-1000 PCL + SDSS  & KiDS-1000 PCL + SDSS  \\ 
        & (PJ-HPD) & (Marginal)  &  (PJ-HPD)  & (Marginal) \\\hline \\[-0.2cm]
        $S_8$             & $\bm{0.754_{-0.029}^{+0.027}}$    & $\bm{0.742_{-0.023}^{+0.034}}$    & $\bm{0.771_{-0.032}^{+0.006}}$     & $\bm{0.762_{-0.023}^{+0.017}}$  \\[0.3cm] 
        $\Omega_{\rm c} h^2$    & $0.064_{-0.013}^{+0.087}$    & $0.091_{-0.034}^{+0.070}$        & $\bm{0.115_{-0.008}^{+0.011}}$     & $\bm{0.114_{-0.009}^{+0.010}}$ \\[0.3cm] 
        $\Omega_{\rm b} h^2$    & $0.022_{-0.001}^{+0.003}$    & $0.021_{-0.001}^{+0.003}$        & $0.022_{-0.003}^{+0.002}$          & $0.024_{-0.003}^{+0.001}$  \\[0.3cm]  
        $h$                     & $0.657_{-0.003}^{+0.118}$    & $0.757_{-0.059}^{+0.049}$        & $\bm{0.664_{-0.022}^{+0.014}}$     & $\bm{0.674_{-0.025}^{+0.010}}$  \\[0.3cm]  
        $n_{\rm s}$                  & $1.000_{-0.074}^{+0.080}$    & $1.000_{-0.080}^{+0.074}$         & $1.003_{-0.073}^{+0.075}$          & $1.050_{-0.118}^{+0.032}$  \\[0.3cm] 
        \hline\\[0.05cm]
        $\Omega_{\rm m}$       & $0.202_{-0.053}^{+0.127}$    & $0.2149_{-0.062}^{+0.125}$        & $\bm{0.312_{-0.018}^{+0.011}}$     & $\bm{0.310_{-0.016}^{+0.015}}$  \\[0.3cm]
        $\sigma_8$             & $0.920_{-0.215}^{+0.159}$    & $0.853_{-0.226}^{+0.103}$         & $\bm{0.756_{-0.044}^{+0.029}}$     & $\bm{0.750_{-0.039}^{+0.037}}$ \\[0.3cm]
        \hline\\[0.05cm]
        $A_{\text{IA}}$   & $\bm{0.396_{-0.941}^{+0.251}}$    & $\bm{0.221_{-0.431}^{+0.639}}$    & $\bm{0.720_{-0.655}^{+0.231}}$     & $\bm{0.471_{-0.440}^{+0.483}}$  \\[0.3cm]  
        $A_{\rm bary}$    & $3.113_{-0.663}^{+0.002}$    & $2.788_{-0.422}^{+0.245}$    & $2.553_{-0.103}^{+0.566}$     & $2.713_{-0.333}^{+0.369}$  \\[0.3cm]  
        $\delta z_1$      & $0.002^{+0.010}_{-0.009}$    & $0.002^{-0.011}_{+0.008}$    & $0.003^{+0.011}_{-0.008}$     & $0.005^{+0.012}_{-0.009}$  \\[0.3cm]  
        $\delta z_2$      & $0.001^{+0.009}_{-0.010}$    & $0.002^{+0.010}_{-0.010}$    & $0.005^{+0.012}_{-0.009}$     & $0.002^{+0.011}_{-0.011}$  \\[0.3cm]  
        $\delta z_3$      & $-0.011^{+0.013}_{-0.005}$   & $-0.016^{+0.012}_{-0.008}$   & $-0.014^{+0.010}_{-0.010}$    & $-0.016^{+0.012}_{-0.009}$ \\[0.3cm]  
        $\delta z_4$      & $-0.013^{+0.008}_{-0.009}$   & $-0.013^{+0.009}_{-0.008}$   & $-0.015^{+0.008}_{-0.008}$    & $-0.014^{+0.008}_{-0.008}$ \\[0.3cm]  
        $\delta z_5$      & $0.006^{+0.009}_{-0.008}$    & $0.009^{+0.007}_{-0.011}$    & $0.002^{+0.006}_{-0.012}$     & $0.007^{+0.009}_{-0.009}$  \\[0.3cm] 
        \hline
    \end{tabular}
    \tablefoot{Here we show the marginalised constraints for using the maximum a posteriori values with the projected joint highest posterior density (PJ-HPD) intervals and the usual one-dimensional marginalised constraints (Marginal). As in \cite{2020-Asgari-2ptsK1000}, we show the constrained parameters in bold, while parameters that are not constrained by the current data or data combination are mostly just reflecting their priors. Although $h$ is constrained by the SDSS data, the value we find is a reflection of our BBN-inspired prior on $\Omega_{\rm b}h^2$.}
\end{table*}

We provide some extra figures and tables in this appendix to complement the relevant information regarding our analysis. Table~\ref{table:1} shows details about the five redshift tomographic bins presented in Fig.~\ref{Fig:Nz} including the m-correction applied to the shear estimates for each tomographic bin. Following that, we present tables and figures related to the Bayesian inference results from Sect. \ref{Sec:Posteriors}. Table \ref{table:FullResults} shows all the model parameters from our \pcl{} analysis (KiDS PCL) and its combination with clustering from SDSS baryon acoustic oscillations and redshift space distortions (KiDS PCL + SDSS) with the constrained parameters shown in bold. We also show both the {maximum a posteriori} results with the projected joint highest probability density (PJ-HPD) and the marginal constraints.

\begin{figure}
  \centering
  \includegraphics[width=0.95\columnwidth]{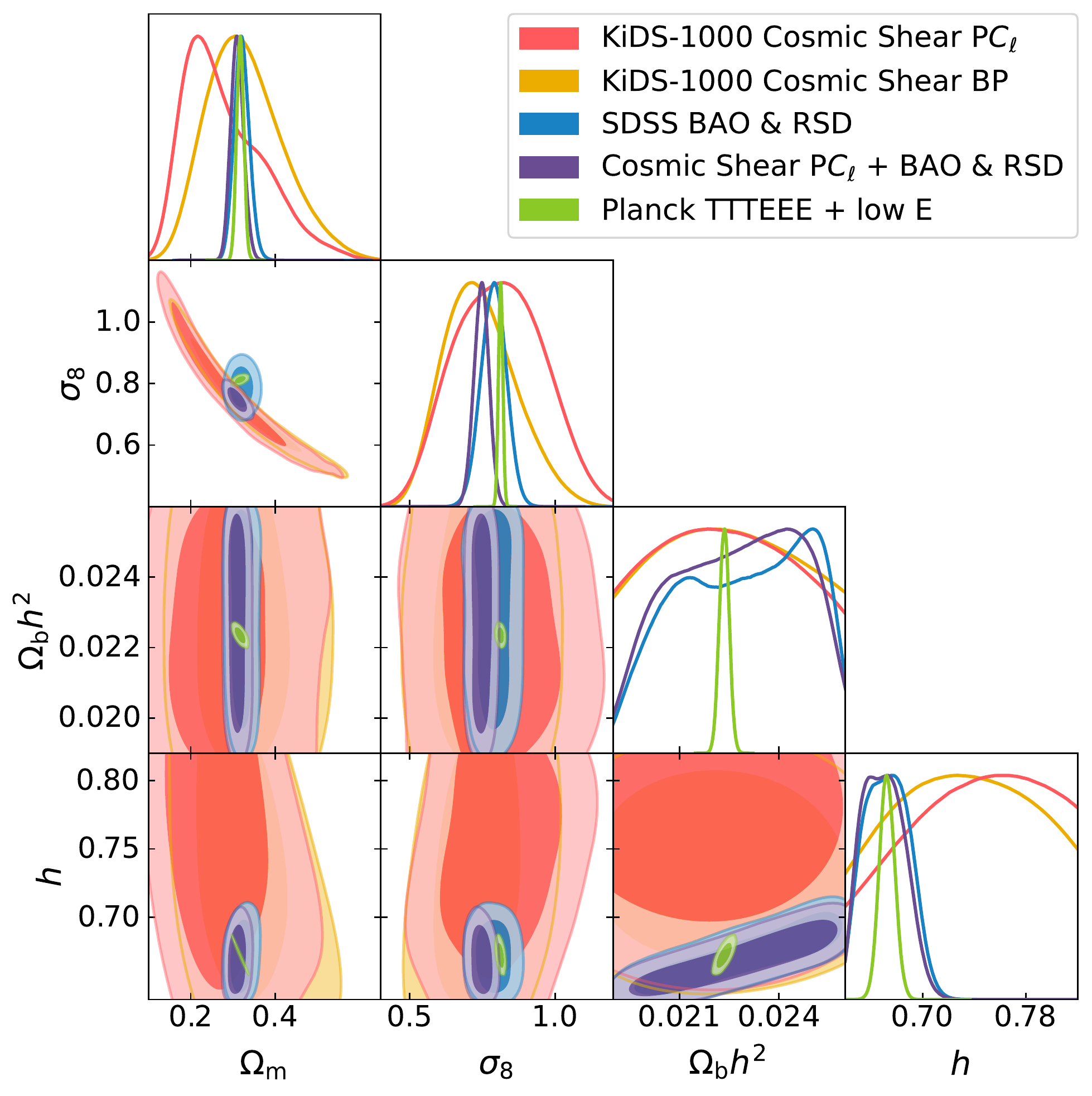}
      \caption{Marginalised 1D and 2D posterior distributions for 68\% (darker) and 95\% (lighter) C.I. in the $\Omega_{\rm m}$, $\sigma_8$, $\Omega_{\rm b}h^2$, and $h$ plane. The degeneracy broken by the introduction of clustering data to the cosmic shear constraints can be understood as better constraining power from the combination with clustering in the $h \times \Omega_{\rm b} h^2$ plane, as well as the BBN-inspired prior. These yield Hubble constant values that are consistent with Planck 2018, in line with what was found by \cite{2019-Cuceu-BAOBBN} for constraints containing SDSS BAO and RSD information.}
         \label{Fig:LCDM_sig8_h_Om}
\end{figure}

\begin{figure}
   \centering
   \includegraphics[width=0.95\columnwidth]{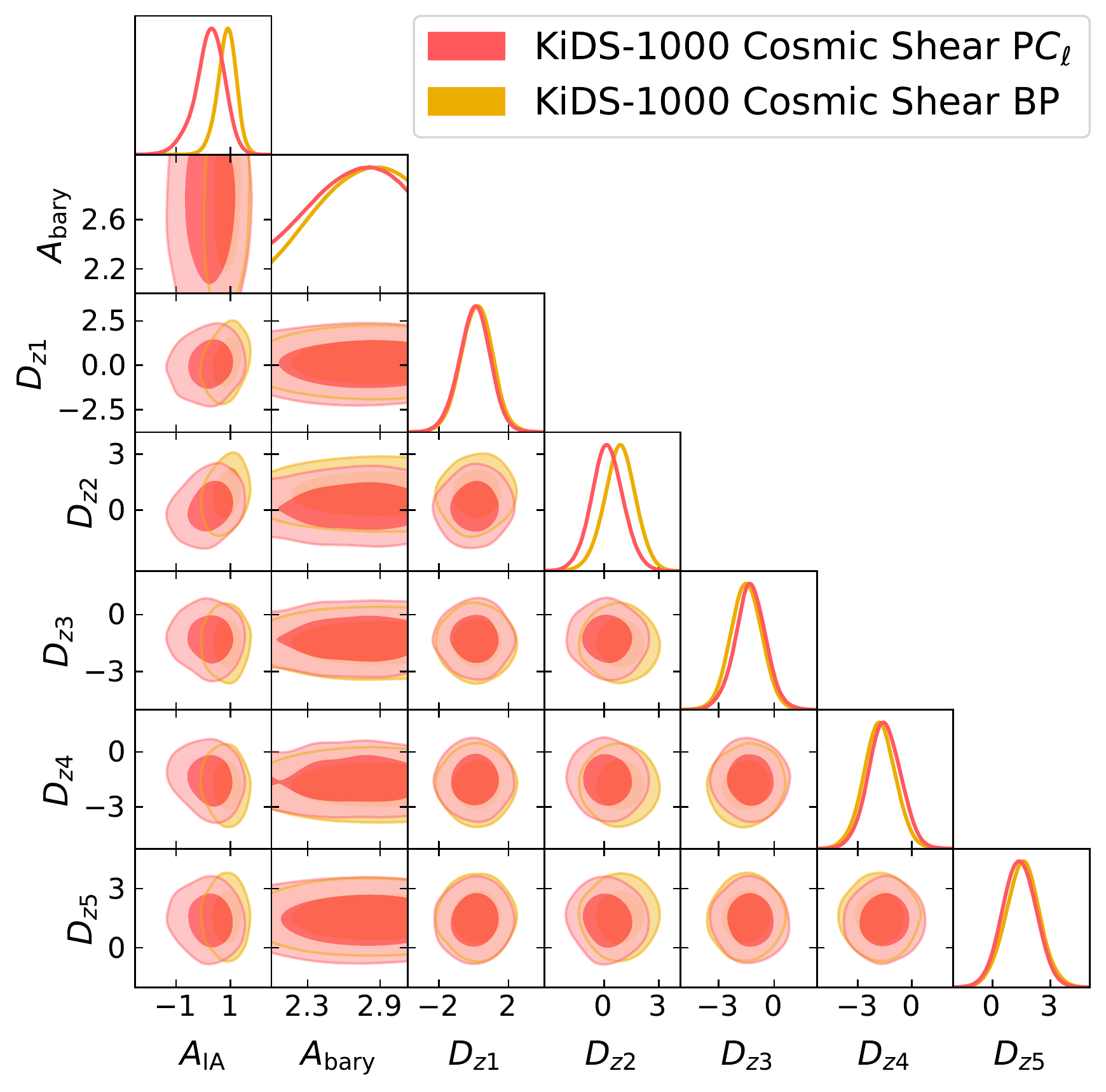}
      \caption{Marginalised 1D and 2D posterior distributions for 68\% (darker) and 95\% (lighter) C.I. in the nuisance parameter space, comparing our \pcl{} results (red) with the band-powers constraints (orange) from \cite{2020-Asgari-2ptsK1000}. The $D_{z_i}$ parameters are shown here as they are sampled, de-correlated displacements calculated from a linear transformation on $\delta_{z_i}$ using the Cholesky decomposition of the redshift covariance, $\tens{C}_{z}$. }
         \label{Fig:LCDM_Redshift}
\end{figure}

\begin{figure}
   \centering
   \includegraphics[width=0.95\columnwidth]{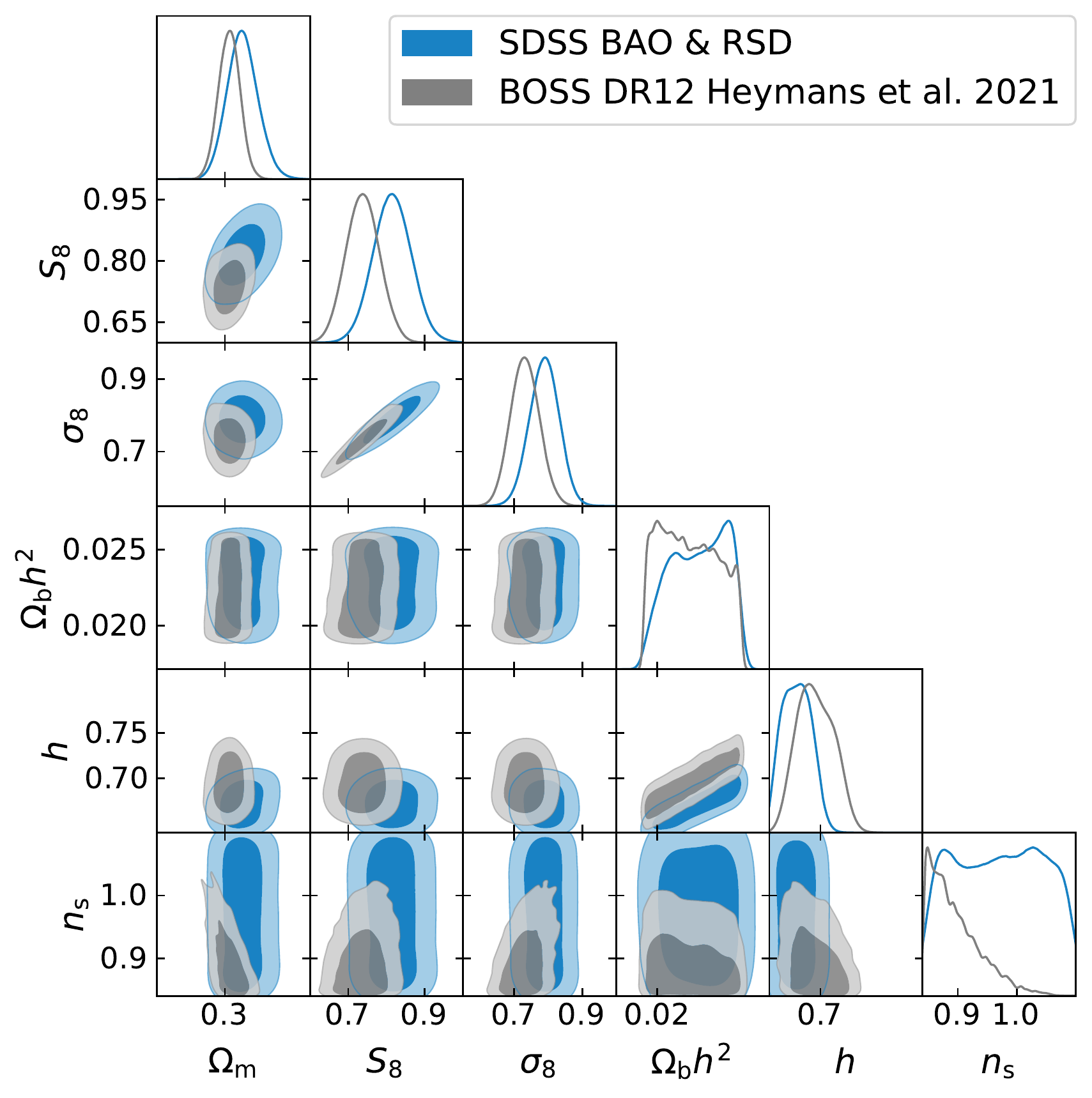}
      \caption{Marginalised 1D and 2D posterior distributions for 68\% (darker) and 95\% (lighter) C.I. in the cosmological parameter space, comparing the results from the  SDSS BAO and RSD (blue) likelihoods we introduce in Sect. \ref{Sec:ExternalData}  with the BOSS DR12 likelihood used in the main KiDS-1000 analysis (grey, \citealt{2017-Sanchez-BOSS}). }
         \label{Fig:LCDM_SDSS}
\end{figure}

When examining a broader parameter sub-space, considering the dimensionless Hubble parameter (Fig. \ref{Fig:LCDM_sig8_h_Om}), we can see that two main factors play a crucial role in breaking the $\sigma_8 - \Omega_{\rm m}$ degeneracy when combining SDSS and KiDS-1000. The BAO and RSD data help constraining significantly the $h - \Omega_{\rm b} h^2$ plane, but there is a second factor playing a part here: the BBN-inspired prior on the baryonic matter density. As was highlighted in Sect.~\ref{Sec:LikePriors}, the 5$\sigma$ prior on $\Omega_{\rm b} h^2$ based on BBN constraints from \cite{2014-Olive-BBN} was a conservative equivalent to what has been considered in \cite{2019-Cuceu-BAOBBN}. However, it is a prior that has an impact when probing  SDSS BAO and RSD data. This has a significant role in constraining the Hubble parameter, which impacts the $h - \Omega_{\rm b} h^2$ plane and consequentially the $h  - \sigma_8$ plane, resulting in tighter constraints when combining clustering data with cosmic shear information. Here, however, we advocate that the inclusion of BBN data would be more constraining than our prior choice \citep{2019-Cuceu-BAOBBN}, therefore raising the point that such a choice was, from this perspective, conservative. 

In Fig. \ref{Fig:LCDM_Redshift} we show a comparison between \pcl{} and band-power measurements \citep{2020-Asgari-2ptsK1000} for the sub-space of nuisance parameters, including the amplitude of intrinsic alignments ($A_{\text{IA}}$), the amplitude of baryonic feedback ($A_{\rm bary}$), and the uncorrelated redshift displacements. In order to properly implement correlated Gaussian priors, we linearly transform the redshift displacements $\delta_{z_i}$ to $D_{z_i}$ using a Cholesky decomposition of the covariance of the redshift distributions. Therefore, the $D_{z_i}$ are uncorrelated. Naturally, this is transformed back into the original correlated space before using the redshift distributions in the forward-modelling described in Sect.~\ref{Sec:Theory}. The results reported by Table \ref{table:FullResults} are for $\delta_{z_i}$, that is the correlated redshifts.
Figure \ref{Fig:LCDM_Redshift} shows good agreement between our analysis and band-powers from \cite{2020-Asgari-2ptsK1000} for all nuisance parameters, except for the amplitude of intrinsic alignments, as discussed in detail in Sect.~\ref{Sec:Posteriors}.

Finally, in Fig. \ref{Fig:LCDM_SDSS} we show a comparison between the SDSS BAO and RSD likelihood and data we present in Sect.~\ref{Sec:ExternalData} with the clustering likelihood used in \cite{2020-Heymans-KiDS1000-Cosmology} from BOSS DR12 \citep{2017-Sanchez-BOSS}. The constraints are in agreement but differ due to containing different data sets and approaches. The SDSS likelihood we use extracts BAO and RSD information from BOSS and eBOSS LRGs, Ly-$\alpha$ auto-correlations and its cross-correlations with quasars. Cosmology is then fitted to BAO and RSD quantities. In contrast to this, the BOSS DR12 approach from \cite{2017-Sanchez-BOSS} used in \cite{2020-Heymans-KiDS1000-Cosmology} and \cite{2020-Tilman-BOSS-LCDM} fits cosmology directly to measurements of the anisotropic galaxy clustering correlation functions.

\section{B-modes systematics analysis}
\label{Apdx:BModesSyst}

\begin{figure}
  \centering
  \includegraphics[width=\columnwidth]{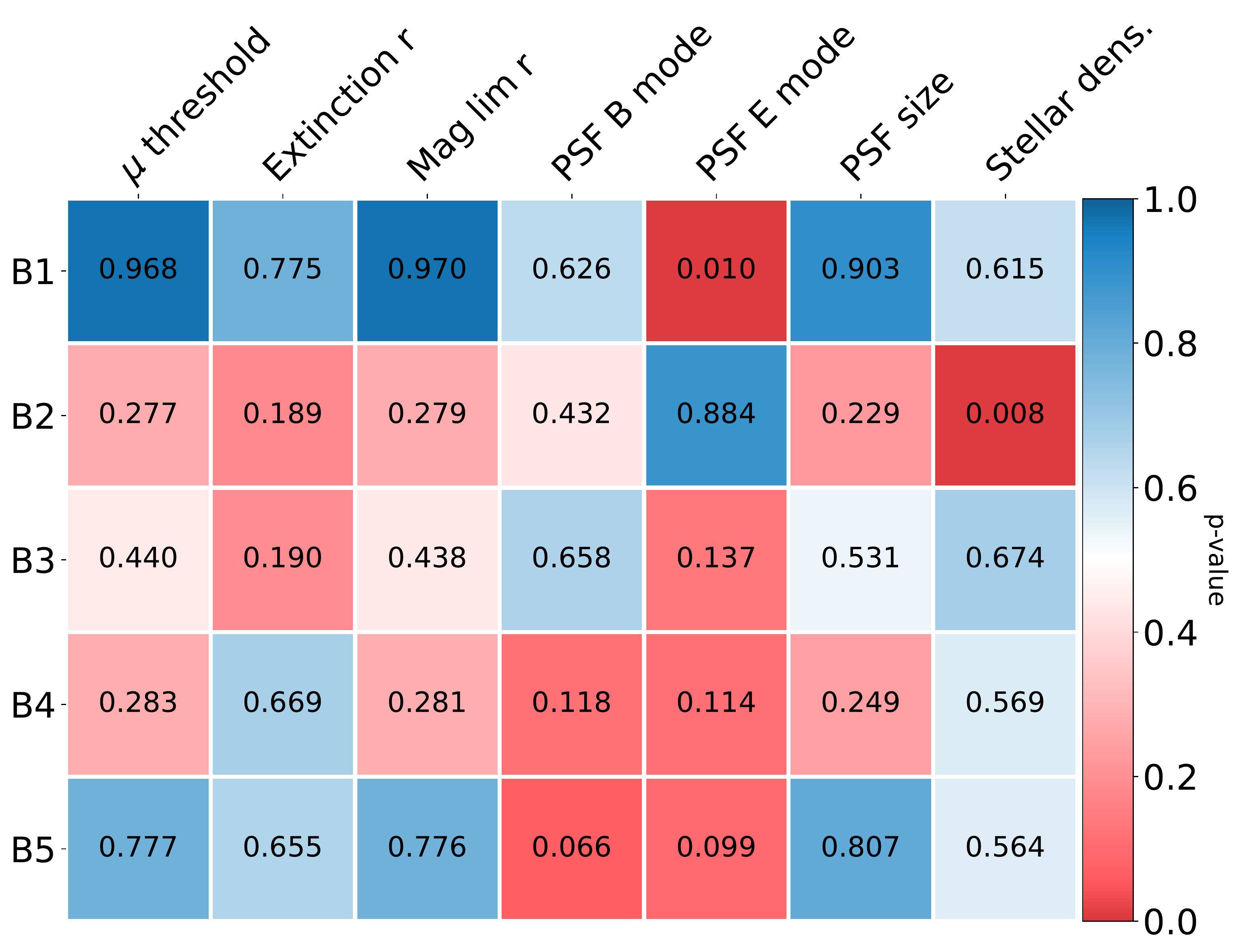}
      \caption{P-values for a null detection from cross-correlations between the measured B-modes and different systematics. The cross-correlations between $B_1$ and PSF E-mode are shown in detail in Fig. \ref{Fig:Syst-PSF-E}, while those between $B_2$ and stellar density are shown in Fig.~\ref{Fig:Syst-Stellar-B}.}
         \label{Fig:Syst-RedChi2-B}
\end{figure}
\begin{figure}
  \centering
  \includegraphics[width=0.8\columnwidth]{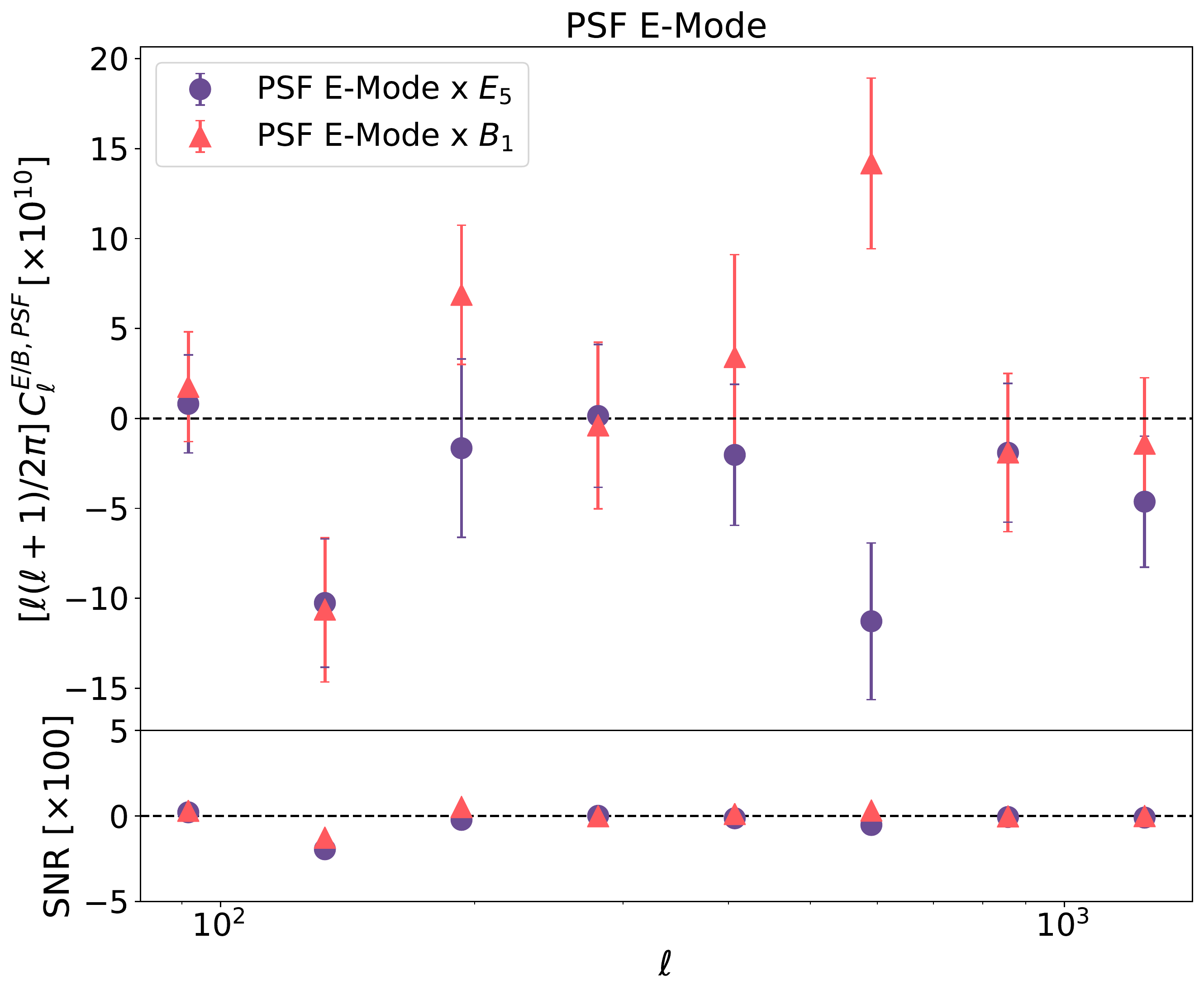}
      \caption{Top panel: Cross-angular power spectra between PSF E-mode component and $E_5$ (purple circles) or $B_1$ (red triangles). Bottom panel:  Signal-to-noise ratio (S/N) as defined by Eq. \eqref{Eq:SNR}.The reduced $\chi^2$ for a null signal is 2.16 (p-value $=0.027$) for $E_5$ and 2.51 (p-value $=0.010$) for $B_1$. The S/N is below $5\times 10^{-2}$ for all the bandwidths, meaning that the estimated data covariance is much larger than this correlation.}
         \label{Fig:Syst-PSF-E}
\end{figure}

\begin{figure}
  \centering
  \includegraphics[width=0.8\columnwidth]{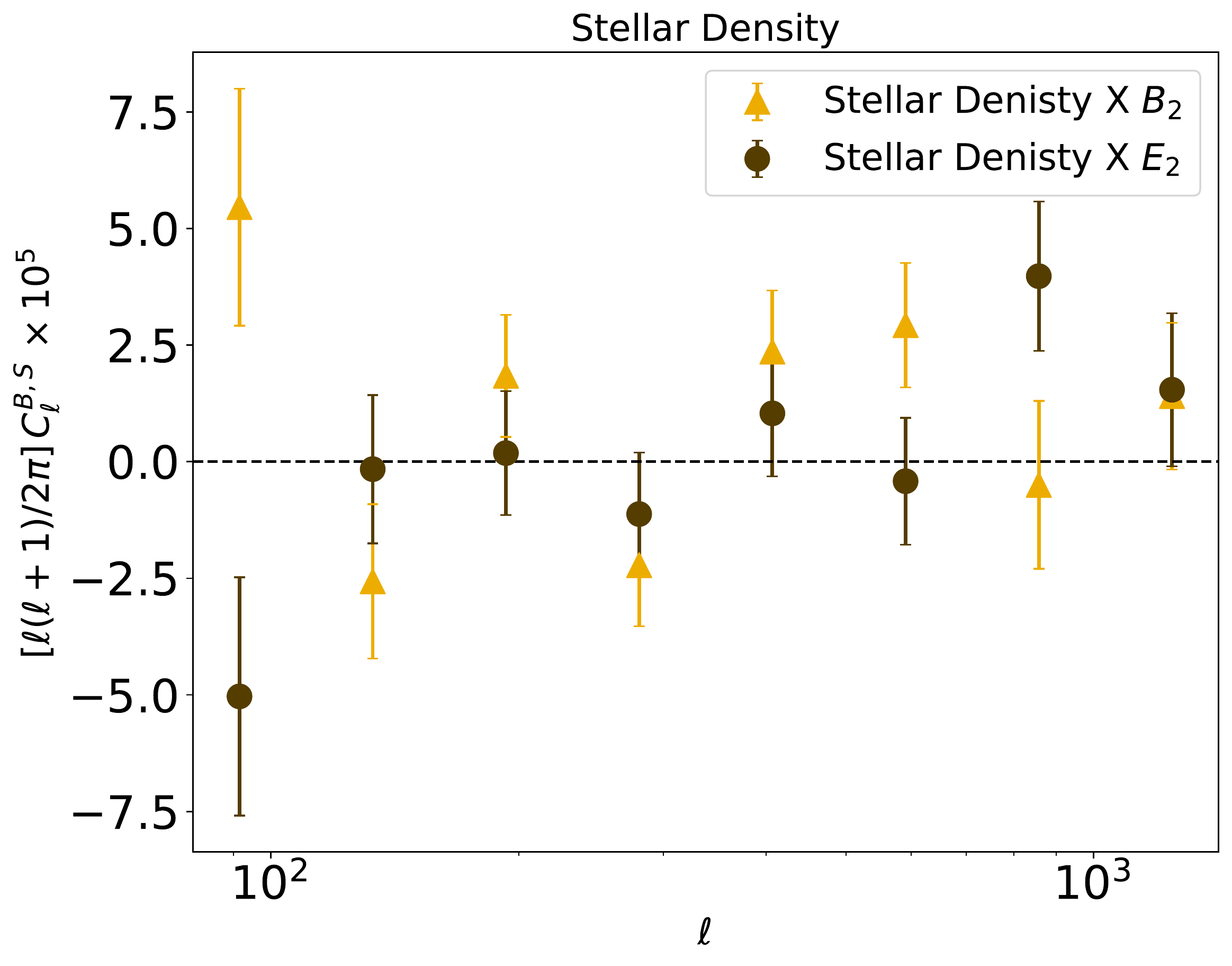}
      \caption{Cross-angular power spectra between stellar density and $E_2$ (dark circles) or $B_2$ (yellow triangles). For these cases the reduced $\chi^2$ for a null signal is 1.54 and 2.60, respectively. }
         \label{Fig:Syst-Stellar-B}
\end{figure}
Here, we present the complete systematic analysis for the B-modes in the data. In the context of cosmic shear in a standard \lcdm{} cosmology, the pure B-modes do not contain cosmological information. This means that any measurements of B-modes that are inconsistent with zero could be a potential red flag for systematic contamination. Partial sky observations cause E/B-mode leakage, where part of the power measured from potential B-modes leaks into the E-modes; such modes should be excluded from the analysis. This could be potentially problematic; however, from the scenario we are seeing in Fig. \ref{Fig:Syst-RedChi2-B}, our B-modes have small correlations with systematics. Figure \ref{Fig:PCL-Measurements} shows that the \pcl{} B-modes are consistent with zero, indicating that any leakage from any significant correlations between B-modes and systematics are unlikely to interfere with the measured E-modes used for our cosmological analysis.

The highest correlations with B-modes manifest in the cross-correlations with the PSF E-mode  and the stellar density from Gaia DR2, with a reduced $\chi^2 = 2.51$ and $2.60$, respectively. For the PSF ellipticities, it is reasonable, however not guaranteed, to assume that the contamination is linear in ellipticity space. Therefore, we can compare these with the errors measured from the mocks (see Sect. \ref{Sec:CovMat}). The signal-to-noise ratio is defined as
\begin{align}
    \text{S/N} = \left(\frac{\tilde{C}^{\text{X}_i,\text{syst}}_{\ell}}{\sigma_{\ell}^{\text{X}_i,\text{X}_i}}\right)\, ,
    \label{Eq:SNR}
\end{align}
where $\sigma_{\ell}^{\text{X}_i,\text{X}_i}$ is the standard deviation for the auto-power spectra $\text{X}_i$, where $\text{X} = {E, B}$. Figure \ref{Fig:Syst-PSF-E} shows the cross-correlations between PSF E-mode and $B_1$($E_5$), as well as the S/N for these cross-correlations. From the S/N analysis, we can see that even with a relatively high $\chi^2_{\text{red}}$, the cross-correlations signal is subdominant for both {$B_1$ $\times$ PSF-E} and {$E_5$ $\times$ PSF-E} (more details in Sect. \ref{Sec:Systematics}). For most bandwidths, the S/N is very close to zero, with the exception of one with $\text{S/N} \approx 5\times10^{-2}$.

For the case shown in Fig.~\ref{Fig:Syst-Stellar-B},
 we cannot perform a S/N analysis as the units between the $\tilde{C}^{B_2,\text{syst}}_{\ell}$ and the data covariance do not match. It is possible that we are observing a stellar contamination for the second tomographic bin B-mode. Nonetheless, this possible contamination has no impact in the E-modes used in our cosmological analysis. We know that even by considering E/B-mode leakage, the auto and cross angular power spectra with $B_2$ shown in Fig. \ref{Fig:PCL-Measurements} are consistent with zero. Another piece of evidence that this possible contamination does not affect our cosmological signal is the cross-correlation between $E_2$ and stellar overdensity, with a reduced $\chi^2 \approx 1.5$ (p-value $=0.13$), also shown in Fig.~\ref{Fig:Syst-Stellar-B}. Had this potential contamination leaked from the B-modes to the E-modes in the second tomographic bin, we would have observed a higher correlation between $E_2$ and stellar density, which does not seem to be the case. We conclude that any significant correlations between B-modes and systematics do not affect any of our cosmic shear data-vectors in a way that could potentially bias $S_8$ and other cosmological parameters.

\end{document}